\documentclass[pre,aps,twocolumn,superscriptaddress,showpacs]{revtex4}
\usepackage{amssymb,epsfig,amsmath,graphicx,textcomp,float,color,cases,epstopdf}
\usepackage{soul}
\usepackage[colorlinks=true, urlcolor=blue, anchorcolor=blue, citecolor=blue,filecolor=blue,linkcolor=blue,menucolor=blue,pagecolor=blue]{hyperref}

\begin{document}
\title{Fluctuations of absorption of interacting diffusing particles by multiple absorbers}
\author{Tal Agranov}
\email{tal.agranov@mail.huji.ac.il}
\affiliation{Racah Institute of Physics, Hebrew University of Jerusalem, Jerusalem 91904, Israel}
\author{Baruch Meerson}
\email{meerson@mail.huji.ac.il}
\affiliation{Racah Institute of Physics, Hebrew University of Jerusalem, Jerusalem 91904, Israel}
	
		
\begin{abstract}
We study fluctuations of particle absorption by a three-dimensional domain with multiple absorbing
patches. The domain is in contact with a gas of interacting diffusing particles. This problem is
motivated by living cell sensing via multiple receptors distributed over the cell surface.
Employing the macroscopic fluctuation theory, we calculate the covariance matrix
of the particle absorption by different patches, extending previous works which addressed fluctuations of a single current.  We find a condition when the sign of correlations between different patches is fully determined by the transport coefficients
of the gas and is independent of the problem's geometry.  We show that the fluctuating particle flux field typically develops vorticity. We establish a simple connection between
the statistics of particle absorption by all the patches combined and the statistics of current in a non-equilibrium steady state in one dimension. We also discuss
connections between the absorption statistics and (i) statistics of electric currents in multi-terminal diffusive conductors and (ii) statistics of wave transmission through disordered media with multiple absorbers.

\end{abstract}
\maketitle
	
\section{Introduction}
\label{intro}

Fluctuations of currents of matter and energy is an important subject of nonequilibrium statistical mechanics. A prototypical model problem, which has attracted much attention, involves
a diffusive lattice gas driven by two reservoirs of particles kept at different densities \cite{lebowitz,Derrida2007,shpiel,bd,MFTreview}. This simple setting has a direct relevance to experiment in at least two different contexts: statistics of electric current in mesoscopic conductors \cite{jor,meso,multi} and statistics of wave transmission through disordered media \cite{penini,shapiro2}. Here we consider a different but closely related problem: transport of diffusing molecules into the living cell through receptors distributed on its surface \cite{berg,bergbook}. The surrounding gas serves as a finite-density reservoir, whereas the cell receptors can be modeled as  reservoirs kept at zero gas density. In their pioneering 1977 paper, Berg and Purcell \cite{berg} evaluated the expected  steady-state current of non-interacting diffusing particles into a single receptor.  Their motivation was to assess physical limitations of the cell's ability to sense changes in the environmental concentration \cite{berg,bergbook}. Building on their work, here we aim at  (i) evaluating the current \emph{fluctuations} at each receptor during a given time, and (ii) accounting for inter-particle \emph{interactions} in the surrounding gas. These extensions are presently possible due to recent advances in the fluctuating hydrodynamics \cite{Spohn} and the macroscopic fluctuation theory (MFT)  \cite{MFTreview} of diffusive lattice gases. These formalisms have already been successfully used in the simplest two-reservoir setting \cite{Derrida2007,MFTreview,shpiel,bd}, and in several non-stationary settings which involved a single current.  As we will show here, the presence of multiple absorbing patches, leading to multiple currents, allows one to ask new questions, and brings new effects. One new question concerns the joint probability distribution of, and correlations between, the absorption currents at different receptors. One new effect is that the most probable particle flux field, conditioned on a specified joint absorption statistics, exhibits a large-scale vorticity.

We will model the surrounding gas of interacting particles as a diffusive lattice gas \cite{Spohn,Liggett}. The large-scale long-time behavior of such gases can be described by fluctuating hydrodynamics \cite{Spohn,KL}. The average particle density  $\rho(\mathbf{x},t)$ of a lattice gas obeys a diffusion equation
\begin{equation}
	\partial_t \rho = \nabla \cdot \left[D(\rho) \nabla \rho\right] , \label{diff}
\end{equation}	
whereas macroscopic fluctuations are described by the conservative Langevin equation \cite{Spohn,KL}
\begin{equation}
	\partial_t \rho = -\nabla \cdot \mathbf{J},\quad\mathbf{J}=-D(\rho) \nabla \rho-\sqrt{\sigma(\rho)}\boldsymbol{\eta}(\mathbf{x},t),\label{lang}
\end{equation}	
where $\boldsymbol{\eta}(\mathbf{x},t)$ is a zero-mean Gaussian noise, delta-correlated in space and in time. As one can see
from Eq.~(\ref{lang}), a diffusive lattice gas is completely specified by two transport coefficients: the diffusivity $D(\rho)\geq 0$ and the mobility $\sigma(\rho)\geq 0$. For lattice gases $\sigma(0)=0$.
	
The simplest example of a diffusive lattice gas is a gas of non-interacting Random Walkers (RWs). For the RWs one has $D(\rho)=D_0=\text{const}$, and $\sigma(\rho)=2D_0\rho$ \cite{Spohn}. An example of \emph{interacting} lattice gas is the Simple Symmetric Exclusion Process (SSEP). Here at each time step a particle can jump, with equal probability, to any \emph{empty} neighboring site. The average behaviors of the RWs and the SSEP turn out to be identical: they share the same density-independent diffusivity $D_0$. The inter-particle interactions of the SSEP are manifested at the level of fluctuations, as the SSEP's mobility  $\sigma(\rho)=2D_0\rho(1-\rho)$ is a non-linear function of $\rho$ \cite{Spohn}. The lattice gases are not the only systems describable by the Langevin equation (\ref{lang}). Important additional examples describe  transport of noninteracting electrons in mesoscopic materials at zero temperature \cite{jor} and wave transmission in disordered media \cite{penini,shapiro2}.

\begin{figure}
\includegraphics[width=6.5cm]{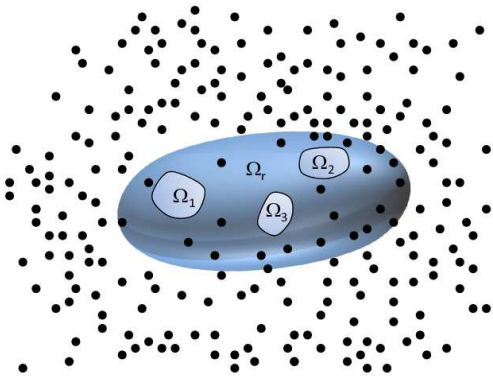}
\caption{A sketch of our system at $t=0$. A gas of particles (black dots) with a constant density surrounds a domain outlined by the thick line. The reflecting part  $\Omega_r$ of the domain boundary is shown in gray, the absorbing patches $\Omega_{1}$, $\Omega_{2}$ and $\Omega_{3}$ are shown in white.}
\label{hure}
\end{figure}

Now we formulate the model which we will study in this paper. Consider a lattice gas of initially uniform density $\rho(\mathbf{x},t=0)=\rho_0$ which fills the whole space outside of a simple-connected three-dimensional domain (the cell) of the characteristic linear size $L$. The domain boundary $\Omega$ includes absorbing patches (the cell receptors) $\Omega_{i},\,i=1,2,\dots,s$. Whenever a particle hits any of these, it is immediately absorbed. Whenever a particle hits the rest of the surface, $\Omega_r$ (the cell wall), it is reflected, see Fig.~\ref{hure}.
At times $T$ much longer than the characteristic diffusion time $L^2/D(\rho_0)$, the system reaches a non-equilibrium steady state. In the steady state the average gas density $\bar{\rho}(\mathbf{x})$ and the
average absorption current $\bar{n}_i=\bar{N}_i/T$ into each absorbing patch are independent of time. Here $\bar{N}_i$ is the average number of particles absorbed by the $i$-th patch during the time interval $0<t<T$. In this work we will determine the joint probability distribution,  ${\mathcal P}(\delta n_1,\delta n_2,\dots,\delta n_s;\rho_0,T)$,  of observing small fluctuations, $\delta n_i=(N_i-\bar{N}_i)/T$, of the absorption currents during the time $T$. As we show here, this multi-variate probability distribution is Gaussian and given by the expression
\begin{equation}\label{mainresult}
     {\mathcal P} \simeq
     \frac{T^{s/2}}{(2\pi)^{s/2}\,{[\text{det}\,\boldsymbol C]^{1/2}}}\exp\left(-\frac{T}{2}\sum_{i,j=1}^{s}\delta n_iC_{ij}^{-1}\delta n_j\right),
\end{equation}
where  $\boldsymbol C$ is an $s\times s$ positive-definite symmetric matrix which depends on $\rho_0$ and on the geometry of the problem, but is independent of time. We obtain a general condition when the \emph{sign} of the currents' cross-correlation $\overline{ \delta n_i \delta n_j }$ is independent of the geometry and completely specified
by the transport coefficients $D(\rho)$ and $\sigma(\rho)$ of the gas.

Further, we show that the cross-correlation between the current into a single absorbing patch and the total current into \emph{all} patches combined has a simple structure where the dependence on the problem's geometry is factorized out. We apply this result to the Berg-Purcell model of a living cell \cite{berg}. An important finding of Ref. \cite{berg} was the dependence of the expected total current on the number of receptors distributed on the cell's surface. We extend their result by finding how the number of receptors affects \emph{correlations}, and also account for interactions.

Our calculations employ the MFT in conjunction with the additivity principle. The latter was
proposed by Bodineau and Derrida \cite{bd} in the context of statistics of current in a one-dimensional lattice gas driven
by two boundaries. We determine the optimal (most probable) spatial profiles of the fluctuating gas density and flux fields, conditioned on the specified
absorption current into each patch. By virtue of the additivity principle the optimal absorption current into the $i$-th patch is independent of time and equal to $n_i=N_i/T$. In order to calculate the variance of the probability distribution (\ref{mainresult}), we use the approach of Ref. \cite{KrMe} and linearize the MFT equations around the deterministic solution. The additivity principle and linearization enable us to solve the problem in quite a general form and for an arbitrary diffusive lattice gas.
We show that the optimal flux field, conditioned on the specified
absorption current into each patch, exhibits a large-scale vortex structure. This feature is unique to multi-reservoir systems sustaining multiple currents, and it appears even when the surrounding gas is modeled as non-interacting RWs. The vorticity is absent
in systems with a single current \cite{shpiel}. Remarkably, it is also absent when the process is conditioned on the total absorption current into \emph{all} patches combined.

Although the original motivation for this work came from a living cell sensing via multiple receptors, many of our results can be generalized to other diffusive systems which are driven by multiple reservoirs and therefore sustain multiple currents. The reservoirs can be disjoint (rather than placed on a common reflecting boundary), and the system can be finite. Still, for concreteness we focus here on the setting shown in Fig. \ref{hure}.

Here is how the remainder of the paper is organized. In Sec. \ref{gena} we present the MFT formulation of the problem. The deterministic limit is discussed in Sec. \ref{m}. The joint distribution of fluctuating absorption currents is obtained in Sec. \ref{solving} and analyzed in Sec. \ref{covv}. In Sec. \ref{compa} we briefly discuss the shot-noise-driven fluctuations of current in multi-terminal diffusive conductors, previously studied in Ref. \cite{multi}. In Sec. \ref{tot} we determine the statistics of  total absorption current into all patches combined, and relate it to previous findings for single-current settings. We also find, in Sec. \ref{corberg}, the cross-correlation between the total current and the current into a single absorbing patch and apply this finding in Sec. \ref{apberg} to the Berg-Purcell model. In Sec. \ref{flu2} we study optimal fluctuations of the density field and of the flux field and uncover a large-scale vortex structure of the optimal flux field. Section \ref{flu2} also discusses the particular case of the density and flux fields conditioned on the total absorption current.  We discuss our main results in Sec. \ref{conc}. Some of the technical derivations are relegated to appendixes.

\section{Macroscopic fluctuation theory of joint absorption statistics}
\label{gena}

The starting point of the MFT formulation is the Langevin equation (\ref{lang}) with the boundary conditions
\begin{eqnarray}\label{blang}
	&&\rho(\mathbf{x}\in{\Omega_i},t)=0,\,i=1,2,\dots,s
\end{eqnarray}
at the absorbing patches, and
\begin{eqnarray}\label{blang2}
	&&\mathbf{J}(\mathbf{x}\in{\Omega_r},t)\cdot\hat{n}=0
\end{eqnarray}
at the reflecting surface. The fluctuating flux field $\mathbf{J}$ is defined in Eq.~(\ref{lang}). Here and in the following $\hat{n}$ denotes a local unit vector normal to the domain boundary
and directed into the domain.  At $t=0$ the gas has a uniform density $\rho_0$,
\begin{equation}\label{initial}
\rho(\mathbf{x},t=0)=\rho_0.
\end{equation}
The boundary condition at infinity is, therefore,
\begin{eqnarray}\label{blang3}
\rho(|\mathbf{x}|\rightarrow\infty,t)=\rho_0.
\end{eqnarray}
In Appendix \ref{mform} we derive the MFT equations for the joint absorption statistics. The derivation, by now pretty standard \cite{MFTreview}, yields the governing equations [Eqs.~(\ref{d11}) and~(\ref{d12}) below] and problem-specific boundary conditions. We repeat the derivation here in order to establish the previously unknown boundary conditions on the absorbing patches. The derivation starts from a path-integral formulation for Eq.~(\ref{lang}) with specified numbers of absorbed particles $N_i$ by time $T$.
The derivation exploits a large parameter -- the typical number of particles in relevant regions of space -- to perform a saddle-point evaluation of the path integral. The ensuing minimization procedure yields the Euler-Lagrange equations which can be cast into a Hamiltonian form for the optimal density history $\rho(\mathbf{x},t)$ (where $\rho$ plays the role of a ``coordinate") and the conjugate momentum density $p(\mathbf{x},t)$:
\begin{eqnarray}
\partial_t \rho &=& \frac{\delta H}{\delta p}= \nabla \cdot \left[D(\rho) \nabla \rho-\sigma(\rho) \nabla p\right], \label{d11} \\
\partial_t p &=& -\frac{\delta H}{\delta \rho}=- D(\rho) \nabla^2 p-\frac{1}{2} \,\sigma^{\prime}(\rho) (\nabla p)^2. \label{d12}
\end{eqnarray}
The Hamiltonian $H$ is given in Eq.~(\ref{hamilton1}), and  the prime denotes the derivative with respect to the single argument. The optimal
flux field is given by:
\begin{equation}\nonumber
\mathbf{J}=-D(\rho) \nabla \rho+\sigma(\rho) \nabla p.
\end{equation}
The boundary conditions in time are
\begin{eqnarray}
\rho(\mathbf{x},t=0)&=&\rho_0,\label{bt0}\\
p(\mathbf{x},t=T)&=&0\label{bt1},
\end{eqnarray}
where here and in the following $\mathbf{x}$ is outside of the domain.
The boundary conditions in space are the following. Far from the domain the gas is unperturbed, so
\begin{eqnarray}
\rho(|\mathbf{x}|\rightarrow\infty,t)&=&\rho_0,\label{inftyq}\\
p(|\mathbf{x}|\rightarrow\infty,t)&=&0.\label{inftyp}
\end{eqnarray}
On the domain boundary we have
\begin{eqnarray}
\rho(\mathbf{x}\in{\Omega_i},t)&=&0,\label{bcgen1}\\
p(\mathbf{x}\in{\Omega_i},t)&=&\lambda_i,\label{bcgen12}\\
\nabla \rho(\mathbf{x}\in{\Omega_r},t)\cdot\hat{n}&=&\nabla p(\mathbf{x}\in{\Omega_r},t)\cdot\hat{n}=0,\label{bcgen13}
\end{eqnarray}	
where $\lambda_i$  are \textit{a priori} unknown Lagrange multipliers which are ultimately set by the $s$ constraints of
having the specified numbers $N_i$ of particles absorbed by the patches by time $T$. The boundary conditions (\ref{bcgen12})
for $p$ generalize their simple analogs in single-current settings \cite{bd,MFTreview,shpiel,main,b,fullabsorb}.

Having solved the coupled nonlinear partial differential equations (\ref{d11}) and (\ref{d12}) with the boundary conditions in space and time, one determines the optimal history of the system conditioned on $N_i$, $i=1,2,\dots, s$.  With the solutions at hand, one can
calculate the action $S$ which yields $-\ln {\mathcal P}$ up to a pre-exponential factor:
\begin{eqnarray}
\label{actionmain}
&-&\ln {\mathcal P}(N_1,N_2,\dots,N_s;\rho_0,T) \nonumber \\
&\simeq&S =\frac{1}{2}\,\int_0^T dt \int d\mathbf{x}\,
\sigma(\rho)\, (\nabla p)^2.
\end{eqnarray}
Here and in the following the volume integral $\int d\mathbf{x}$ is performed over all space outside of the domain.
For typical fluctuations the pre-exponential factor in $\mathcal{P}$, see Eq.~(\ref{mainresult}), is
determined from normalization to unity.

\section{Deterministic theory}\label{m}

The choice $\lambda_i=0$ sets $p$ to vanish at all times and describes the deterministic solution,
where all $N_i$ are equal to their expected values $\bar{N}_i$. In this case Eq.~(\ref{d11})  reduces to the deterministic diffusion equation (\ref{diff}) for the average density. At long times, $T\gg L^2/D(\rho_0)$,  the solution approaches a stationary one, $\bar{\rho}(\mathbf{x})$, obeying the time-independent equation
\begin{eqnarray}
\nabla \cdot \left[D(\bar{\rho}) \nabla \bar{\rho}\right]=0 \label{mfq}
\end{eqnarray}
and the boundary conditions in space. As a result, a steady-state particle flux field, and steady-state currents into the absorbing patches, set in. Essentially, this is the problem which Berg and Purcell \cite{berg} solved for gases with constant diffusivity $D_0$. They did it using an illuminating electrostatic analogy \cite{shpitzer,Rednerbook} which can be easily generalized to a density-dependent diffusivity. Let us introduce the following function of the steady-state density $\bar\rho(\mathbf{x})$:
\begin{equation}\label{phi}
\phi\left[\bar{\rho}\left(\mathbf{x}\right)\right]=\int_{\bar{\rho}(\mathbf{x})}^{\rho_0}D(w)dw.
\end{equation}
Then Eq.~(\ref{mfq}) for $\bar{\rho}(\mathbf{x})$  becomes the Laplace's equation for $\phi$:
\begin{equation}
\nabla^2\phi=0.
\end{equation}
The boundary conditions for $\bar{\rho}(\mathbf{x})$ transform to the following boundary
conditions for $\phi$:
\begin{eqnarray} \label{v}
&&\phi(\mathbf{x}\in{\Omega_i})=V(\rho_0)\equiv\int_{0}^{\rho_0}D(w)dw
,\\
&&\nabla\phi(\mathbf{x}\in{\Omega_r})\cdot\hat{n}=0,\quad\phi(\mathbf{x}\rightarrow\infty)= 0.
\end{eqnarray}
As in Ref. \cite{berg}, $\phi$ can be interpreted as the electrostatic potential outside an insulating domain with boundary $\Omega$ over which $s$ conducting patches $\Omega_i$, held at voltage $V({\rho_0})$, are distributed. Having solved for this potential, one obtains the complete solution of the deterministic problem:  $\bar{\rho}$ is obtained by inverting the relation for $\phi\left(\bar{\rho}\right)$ in Eq.~(\ref{phi}).
For gases with constant diffusivity $D_0$, such as the RWs and SSEP, Eq.~(\ref{phi}) defines a linear relation
\begin{equation}
\phi \left(\bar{\rho}\right)=D_0\left(\rho_0-\bar{\rho}\right)\label{phirw}
\end{equation}
which we will use in the following. The deterministic steady-state flux field $\bar{\mathbf{J}}$ is minus the \emph{electric field} of this system:
\begin{equation}
\bar{\mathbf{J}}=-D(\bar{\rho})\nabla\bar{\rho}=\nabla\phi.\label{j}
\end{equation}
In their turn, the average steady-state currents $\bar{n}_i=\bar{N}_i/T$ are (up to a factor of $4\pi$) the \emph{electric charges} accumulated on the conducting patches:
\begin{equation}
\bar{n}_i=\oint_{\Omega_i}\bar{\mathbf{J}}\cdot\hat{n} \,dS=\oint_{\Omega_i}\nabla\phi\cdot\hat{n} \,dS,\quad i=1,2,\dots, s. \label{current}
\end{equation}
These charges are linearly related to the voltage $V(\rho_0)$ via an $s\times s$ symmetric \emph{capacitance matrix} (which we will denote by $\mathbf A$), determined solely by the problem's geometry. The $i,j$ element of $\mathbf A$ is the charge on the patch $i$ induced by the unit voltage applied to the patch $j$, the rest of the patches being grounded \cite{smythe}. $\mathbf A$ can be expressed via a set of $s$ characteristic electrostatic potentials $\phi_i(\mathbf{x})$. Each of them appears when the corresponding conducting patch $\Omega _i$ is held at unit voltage, the rest of the conducting patches are grounded, and the Neumann boundary condition is specified at the reflecting part of the boundary $\Omega_r$. The capacitance matrix is given by
\begin{equation}\label{a}
A_{ij}=\oint_{\Omega_i} \nabla \phi_j\cdot\hat{n}dS .
\end{equation}
In its turn, $\phi(\mathbf{x})$ from Eq.~(\ref{phi}) can be expressed as
\begin{equation}
\phi(\mathbf{x})=V(\rho_0)\sum_{i=1}^s\phi_i(\mathbf{x}).\label{phii}
\end{equation}
Using Eqs.~(\ref{current})-(\ref{phii}), we obtain
\begin{equation}\label{barn}
\bar{n}_i=V(\rho_0) \sum_{j=1}^{s}A_{ij}.
\end{equation}
To highlight  the symmetry of the capacitance matrix $\mathbf{A}$, let us rewrite Eq.~(\ref{a}) as
\begin{eqnarray}\nonumber\label{a2}
A_{ij}=\oint_{\Omega}\nonumber \phi_i\nabla\phi_j\cdot\hat{n}dS&=&\int d\mathbf{x}\nabla \cdot \left(\phi_i\nabla\phi_j\right)\\ &=&\int d\mathbf{x}\nabla\phi_i\cdot\nabla\phi_j,
\end{eqnarray}
where we used the boundary conditions for $\phi_i$, the Gauss theorem, and the fact  that $\phi_i$ are harmonic functions in the bulk. We will call the integrand of the last expression in Eq.~(\ref{a2}), $\nabla\phi_i\cdot\nabla\phi_j$, \emph{the capacitance density} of the system.
Finally, the total absorption current into all patches in this interpretation is equal to the total charge on all the patches when they are held at voltage $V(\rho_0)$:
\begin{equation}\label{current2}
\bar{n}=\sum_{i=1}^s\bar{n}_i=V(\rho_0)A,
\end{equation}
where
\begin{equation}\label{totalcapacitance}
A\equiv\sum_{i,j=1}^{s}A_{ij}
\end{equation}
is the total capacitance of the system, that is the total charge accumulated on all the patches when they
are kept at unit voltage.

\section{Absorption Statistics}
\label{flu}

\subsection{Solving linearized MFT equations}
\label{solving}

As already mentioned,  we will solve the MFT problem under two simplifying assumptions. The first is the additivity principle \cite{bd}: Being interested in the long-time limit, $T\gg L^2/D(\rho_0)$, we look for \emph{stationary} solutions of Eqs.~(\ref{d11}) and (\ref{d12}) which
satisfy the boundary conditions~(\ref{inftyq})-(\ref{bcgen13}) \cite{blayers}.

The second assumption involves linearization of the stationary MFT equations around the steady-state deterministic solution  $\rho=\bar{\rho}(\mathbf{x})$ and $p=0$. This corresponds to typical, small fluctuations and suffices for the evaluation of the variance of the
joint probability distribution \cite{KrMe}. Note that, for the typical fluctuations, the additivity principle appears to be a safe assumption for all diffusive lattice gases \cite{Bertini2005,BD2005,Hurtado,ZM2016}.

Let us denote small deviations of $\rho$ and $p$ from their average values as $\rho_1(\mathbf{x})=\rho(\mathbf{x})-\bar{\rho}(\mathbf{x})$ and $p_1(\mathbf{x})=p(\mathbf{x})$. They determine time-independent current deviations $\delta n_i=n_i-\bar{n}_i$.
The linearized steady-state MFT equations read:
\begin{eqnarray}
\nabla \cdot\boldsymbol{\delta J}&=&0,\quad\boldsymbol{\delta J}=-\nabla\left[D(\bar{\rho})\rho_1\right]+\sigma(\bar{\rho})\nabla p_1,\label{qstatglin}\\
\nabla^2p_1&=&0 \label{pstatglin}.
\end{eqnarray}
The boundary conditions for $\rho_1$ and $p_1$  follow from Eqs.~(\ref{inftyq})-(\ref{bcgen13}):
\begin{eqnarray}
\rho_1(\mathbf{x}\in{\Omega_i})&=&0,\label{dbcgen1}\\
p_1(\mathbf{x}\in{\Omega_i})&=&\lambda_i,\label{dbcgen12}\\
\nabla \rho_1(\mathbf{x}\in{\Omega_r})\cdot\hat{n}&=&\nabla p_1(\mathbf{x}\in{\Omega_r})\cdot\hat{n}=0\label{dbcgen13},\\
\rho_1(|\mathbf{x}|\rightarrow\infty)&=&
p_1(|\mathbf{x}|\rightarrow\infty)=0.\label{dbcgen14}
\end{eqnarray}	
As one can see, the Laplace's equation (\ref{pstatglin}) for $p_1$ is decoupled from Eq.~(\ref{qstatglin}). Subject to the boundary conditions (\ref{dbcgen12})-(\ref{dbcgen14}), it has a unique solution which can be expressed in terms of the auxiliary potentials $\phi_i(\mathbf{x})$, introduced in the previous section:
\begin{equation}
p_1=\sum_{i=1}^s\lambda_i\phi_i.\label{p1}
\end{equation}
This solution suffices for determining the action (\ref{actionmain}) in terms of $\lambda_i$-s. Indeed, in the leading order Eq.~(\ref{actionmain}) yields
\begin{equation}
\label{actionmainlin}
-\ln {\mathcal P} \simeq S =\frac{T}{2}\,\int d\mathbf{x}\,
\sigma(\bar{\rho})\, (\nabla p_1)^2.
\end{equation}
Plugging Eq.~(\ref{p1}) into Eq.~(\ref{actionmainlin}), we obtain the action in terms of
a bilinear form in $\lambda_i$-s:
\begin{equation}
\label{actionmainlin1}
S =\frac{T}{2}\,\sum_{i,j=1}^s\lambda_i\lambda_j\int d\mathbf{x}\,\sigma(\bar{\rho})\nabla\phi_i\cdot\nabla\phi_j=\frac{T}{2}\boldsymbol{\Lambda}^T\cdot\boldsymbol C\cdot\boldsymbol{\Lambda},
\end{equation}
where $\boldsymbol{\Lambda}$ is a vector with components $\lambda_i$,
and  $\boldsymbol C$ is a $s\times s$ symmetric matrix,
\begin{equation}
C_{ij}= \int d\mathbf{x} \,\sigma (\bar{\rho})\nabla\phi_i\cdot\nabla\phi_j,\label{mat}
\end{equation}
given in terms of the volume integral of the product of the  capacitance density of the system  $\nabla\phi_i\cdot\nabla\phi_j$ and the effective local noise magnitude $\sigma[\bar{\rho}(\mathbf{x})]$. As we will see shortly, this is the \emph{covariance matrix}. It
is fully determined by the deterministic solution, and it plays a crucial role in our results.

What is left is to express the $\lambda_i$-s in Eq.~(\ref{actionmainlin1}) in terms of the current  deviations $\delta n_i$.
This requires solving Eq.~(\ref{qstatglin}) which, once $p_1$ is known, is a Poisson's equation for $\rho_1$:
\begin{equation}
\nabla^2\left[D(\bar{\rho})\rho_1\right]=\nabla\cdot\left[\sigma(\bar{\rho})\nabla p_1\right]=\sum_{i=1}^s\lambda_i\nabla\cdot\left[\sigma(\bar{\rho})\nabla\phi_i\right].\label{pois}
\end{equation}
In Appendix \ref{qrws} we present an explicit solution of this equation for the RWs. For a general lattice gas an explicit solution is unavailable. Still,
we were able to derive an explicit relation for  $\delta n_i$ vs.
$\lambda_i$ by using a  Green's function identity,
see Appendix \ref{lamcur}.
This relation is given by the same matrix $\boldsymbol C$ defined in Eq.~(\ref{mat}):
\begin{equation}
\boldsymbol\delta \mathbf{n}=\boldsymbol C\cdot\boldsymbol\Lambda,\label{sigmalin}
\end{equation}
where $\boldsymbol\delta \mathbf{n}$ is the vector with components $\delta n_i$.
As shown in Appendix \ref{mat1}, the symmetric matrix $\boldsymbol C$ is positive definite
and therefore invertible, enabling one to solve Eq.~(\ref{sigmalin}) for $\boldsymbol\Lambda$. Plugging this inverse relation in the bilinear form (\ref{actionmainlin1}) and using Eq.~(\ref{actionmainlin}), we obtain
\begin{equation}\label{lin}
-\ln {\mathcal P}(\delta n_1,\delta n_2,\dots,\delta n_s;\rho_0,T) \simeq \frac{T}{2}\boldsymbol{\delta n}^T\cdot\boldsymbol C^{-1}\cdot\boldsymbol {\delta n}.
\end{equation}
This probability distribution describes Gaussian fluctuations.   Normalizing it to unity, we arrive at the announced result~(\ref{mainresult}).

\subsection{Covariance matrix}\label{covv}

As is clear from Eq.~(\ref{lin}), the joint statistics of absorption is encoded in
the covariance matrix $\boldsymbol{C}$. The diagonal elements of $\boldsymbol C$
describe the variance of the current into the patch $i$:
\begin{equation}\label{vari}
\overline{\delta n_i^2} = \frac{C_{ii}}{T},
\end{equation}
where the over-line denote averaging with respect to the Gaussian distribution (\ref{mainresult}).
The off-diagonal elements of $\boldsymbol C$ describe cross-correlations between the currents into different patches:
\begin{equation}\label{corr}
\overline{ \delta n_i\delta n_j } = \frac{C_{ij}}{T}.
\end{equation}
A necessary condition for nonzero cross-correlations is inter-particle interactions. This can be seen from an alternative expression for $\boldsymbol C$ which we derived in Appendix \ref{offdiag1}:
\begin{equation}
C_{ij}=\frac{\sigma^{\prime}(0)}{2D(0)}\bar{n}_i\delta_{i,j} + \frac{1}{2} \int
d\mathbf{x}\,\phi_i\phi_j\frac{\left(\nabla\phi\right)^2}{D(\bar{\rho})}
\left[\frac{\sigma^{\prime}\left(\bar{\rho}\right)}{D(\bar{\rho})}\right]^{\prime},\label{offdiag}
\end{equation}
where  $\delta_{i,j}$ is the Kronecker delta. For the non-interacting RWs one has $\sigma^{\prime}(\rho)/D(\rho)=2$, so the integrand vanishes, and one is left with $C_{ij}=\bar{n}_i\delta_{i,j}$,  no correlations. From Eq.~(\ref{vari}) one can see that, for the RWs, the variance of the number of absorbed particles
in each patch, $\overline{\delta N_i^2 }$, is equal to the mean value $\overline{N}_i$, regardless of the system's geometry. In fact, for the RWs the steady-state MFT problem can be solved exactly, without linearization. We performed these calculations and found the complete joint distribution of the number of
absorbed particles. As to be expected, this distribution is equal to a product of independent Poisson distributions with the mean $\overline{N}_i$.

A simple \emph{interacting} lattice gas with zero cross-correlations of absorption by different patches is the zero-range process, where a particle can hop to a neighboring site with a rate
which increases with the number of particles on the departure site but is independent of the number of particle
on the target site \cite{zrp}.  For the zero-range process one has, in the hydrodynamic limit, $D(\rho) = (1/2)\, \sigma^{\prime}(\rho)$ \cite{void}, and the integrand in Eq.~(\ref{offdiag}) vanishes.

Going back to general $D(\rho)$ and $\sigma(\rho)$, we notice that the expression $\phi_i\phi_j\left(\nabla\phi\right)^2/D(\bar{\rho})$ under the integral in Eq.~(\ref{offdiag}) is everywhere positive \cite{decay}. Therefore, if
\begin{equation}
\left[\frac{\sigma^{\prime}\left(\bar{\rho}\right)}{D(\bar{\rho})}\right]^{\prime}<0, \quad \mbox{or} \quad
D(\bar{\rho})\sigma^{\prime\prime}(\bar{\rho})<D^{\prime}(\bar{\rho})\sigma^{\prime}(\bar{\rho}), \label{condition}
\end{equation}
for any value of $\bar{\rho}\in[0,\rho_0]$,
then the currents into different patches $i\neq j$ are all anti-correlated, $\overline{ \delta n_i\delta n_j }<0$, regardless of the system's geometry. In particular, this is true for the SSEP, where $\left[\sigma^{\prime}\left(\rho\right)/D(\rho)\right]^{\prime}=-4$.

If the inequality in Eq.~(\ref{condition}) is reversed, the absorption currents are all positively correlated. This happens for the Kipnis-Marchioro-Presutti (KMP) model \cite{KMP,MFTreview} which describes an ensemble of agents on a lattice which randomly redistribute energy among neighbors. Using the KMP model one can study fluctuations of \emph{energy} absorption by absorbing patches located on the domain boundary. For the KMP model $D(\rho) = D_0 = \text{const}$ and $\sigma\left(\rho\right) = 2D_0\rho^2$, and so $\left[\sigma^{\prime}\left(\rho\right)/D(\rho)\right]^{\prime}=4$.

Remarkably, the same inequality (\ref{condition}) guarantees the validity of the additivity principle for arbitrary currents \cite{ber}, and also determines the sign of the two-point \emph{density} correlation function \cite{bertinistat,tridib}, in single-current systems.

\subsection{A comparison with Sukhorukov and Loss \cite{multi}}
\label{compa}

Sukhorukov and Loss \cite{multi} studied shot-noise-driven fluctuations of current in multi-terminal diffusive conductors.
Although theirs and our geometries are different, their expression~(3.16) for the zero-frequency mode of the power spectrum of correlations of the current has a mathematical structure which resembles that of our covariance matrix (\ref{mat}). In their case it is a volume integral over the  ``conductance density" -- an analog of our capacitance density -- times the local noise magnitude. If we closely examine the governing equations of both systems, this resemblance should not come as a surprise.
Sukhorukov and Loss started from a Boltzmann-Langevin description of the distribution function of electrons (where noise comes from electron scattering on static impurities). Then,  applying a diffusion approximation,
they derived a time-independent linear Langevin equation for the electrical potential  $V(\mathbf{x})$ inside the conductor:  Eqs.~(2.12) and (2.13) of Ref. \cite{multi}. For an isotropic medium their equation can be written as
\begin{equation}\label{multilan}
\nabla\cdot\left[D(\mathbf{x})\nabla V+\sqrt{D\left(\mathbf{x}\right)\Pi\left(\mathbf{x}\right)}\,\boldsymbol{\eta}\right]=0,
\end{equation}
where $\boldsymbol{\eta}$ is a Gaussian noise, delta-correlated in $\mathbf{x}$ and $t$. The transport coefficients $D(\mathbf{x})$ and $\Pi\left(\mathbf{x}\right)$ depend on $\mathbf{x}$ but are independent of the fluctuating potential $V(\mathbf{x})$ and of time.
Equation (\ref{multilan}), therefore, is very different from the nonlinear time-dependent Langevin equation~(\ref{lang}) which was the starting point of our analysis. However, in order to make a progress within the MFT formalism, we employed the additivity principle and linearization. In fact, we could have derived the
same  final results from the following Langevin equation which describes
small quasi-stationary density fluctuations $\rho_1(\mathbf{x})$ around the steady-state average density profile
$\bar{\rho}(\mathbf{x})$:
\begin{equation}\label{linlan}
\nabla \cdot \left\{D\left[\bar{\rho}(\mathbf{x})\right]\nabla\rho_1+D^{\prime}\left[\bar{\rho}(\mathbf{x})\right]
\rho_1\nabla\bar{\rho}+\sqrt{\sigma\left[\bar{\rho}(\mathbf{x})\right]}\,\boldsymbol{\eta}\right\}=0.
\end{equation}
The mathematical difference between the linear Langevin equations~(\ref{multilan}) and (\ref{linlan}) comes from the fact that  the  transport coefficients $D(\rho)$ and $\sigma(\rho)$ in Eq.~(\ref{linlan}) are $\rho$-dependent. In particular, the $\rho$-dependence of $D$ causes an additional contribution to the fluctuating flux field, $D^{\prime}\left(\bar{\rho}\right)\rho_1\nabla\bar{\rho}$, which is  absent in Eq.~(\ref{multilan}). Of course, this difference reflects different physical problems which Ref. \cite{multi} and this work address.

Notwithstanding these differences, there is a close similarity between the two problems. An extensively discussed regime in Ref. \cite{multi} is that of zero-temperature elastic scattering, manifested in a particular form of the spatial mobility profile $\Pi(\mathbf{x})$ in Eq.~(\ref{multilan}). An important finding of Ref. \cite{multi} for this case is the strictly negative sign of the cross-correlations of the current, regardless of the system's geometry. Remarkably, the profile $\Pi(\mathbf{x})$ for such a system turns out to be equivalent, up to irrelevant factors, to the mobility coefficient $\sigma\left[\bar{\rho}(\mathbf{x})\right]$ for the SSEP.  As we saw in the previous section, the cross-correlations of the absorption currents for the SSEP are indeed strictly negative.

Here we deal with arbitrary $D(\rho)$ and $\sigma(\rho)$, where the currents' cross-correlations can behave differently. In particular, the sign of the cross-correlations can be strictly positive, as it happens for the KMP model. In addition, the MFT formalism, which we employ here, makes it possible to determine the optimal fluctuating profiles, as we report in Sec. \ref{flu2}.

\subsection{Statistics of total absorption current}\label{tot}

Given the joint absorption statistics, Eq.~(\ref{mainresult}) or (\ref{lin}), what is the probability density that the \emph{total} current into all the patches combined, $\sum_{i=1}^s{n_i}$, is equal to a prescribed value $n$? To answer this question, we can minimize the action (\ref{lin}) under the constraint $\sum_{i=1}^sn_i=n$ or, equivalently, $\sum_{i=1}^s\delta n_i=n-\bar{n}$. The minimization, performed in Appendix \ref{optimum}, yields the following optimal values of the individual currents $n_i$:
\begin{eqnarray}
n_i&=&\bar{n}_i\frac{n}{\bar{n}}.\label{optn}
\end{eqnarray}
The corresponding Lagrange multipliers $\lambda_i$-s [see Eq.~(\ref{bcgen12})] turn out to be all equal:
\begin{eqnarray}
\lambda_i&=&\lambda\equiv\frac{\delta n}{\bar{n}\alpha(\rho_0)},\label{optlam}
\end{eqnarray}
where
\begin{eqnarray}
\alpha(\rho_0)= \frac{I_2(\rho_0)}{I_1^2(\rho_0)}, \quad I_k(\rho_0)\equiv\int_0^{\rho_0}D(w)\sigma(w)^{k-1}dw.\label{alpha}
\end{eqnarray}
Note that $I_1(\rho_0)=V(\rho_0)$, see Eq.~(\ref{v}). Now we can compute the action by using either Eqs.~(\ref{lin}) and~(\ref{optn}) or Eqs.~(\ref{actionmainlin1}) and~(\ref{optlam}). An explicit result can be obtained with the help of the relation
\begin{equation}\label{cij}
\sum_{j=1}^{s}C_{ij} =  \frac{\bar{n}_i I_2(\rho_0)}{I_1^2(\rho_0)},
\end{equation}
derived in Appendix \ref{optimum}. In this way we arrive at a  Gaussian distribution of the total absorption current $n$:
\begin{equation}
-\ln {\mathcal P}(\delta n;\rho_0,T) \simeq S =
\frac{T\delta n^2}{2\bar{n}\alpha(\rho_0)}.\label{sn}
\end{equation}
The first two distribution cumulants of the number of absorbed particles $N=Tn$, following from Eq.~(\ref{sn}), are
\begin{equation}\label{com}
\frac{\overline {N}}{T}=AI_1(\rho_0)\quad;\quad\frac{\overline{ N^2 }-\overline{ N }^2}{T}=A\frac{I_2(\rho_0)}{I_1(\rho_0)},
\end{equation}
where we have used Eq.~(\ref{current2}). Remarkably, these cumulants are equal to the total capacitance $A$ multiplied by the cumulants of the integrated current, obtained for a \emph{one-dimensional} lattice gas driven by two reservoirs, at $\rho_a=\rho_0$ and $\rho_b=0$, see Eq.~(3) of Ref. \cite{bd}.

Similar relations were established, for \emph{all} cumulants of the current,  in Ref. \cite{shpiel} which dealt
with the SSEP driven by two reservoirs in a
finite system in any spatial dimension. There too the cumulants of the current are equal to the corresponding cumulants of the one-dimensional system multiplied by a constant geometrical factor \cite{shpiel}. As was noticed in Ref. \cite{MFTreview}, this geometric factor is the electric capacitance.

Our setting involves an infinite system with a single reservoir at infinity. Still, as we have just shown,
the first two cumulants of the \emph{total} absorption current  have the same structure as those in the finite two-reservoir settings. (For a spherically symmetric absorber this property was established, for the complete
absorption statistics, in Ref. \cite{fullabsorb}.)
Extending the results of Refs. \cite{shpiel} and \cite{fullabsorb}, the relations (\ref{com}) show that the dependence on the problem's geometry, through the electric capacitance $A$,
is factorized out from the expressions for the current's mean and variance. We now show that the same feature also holds for the
cross-correlations.

\subsubsection{Cross-correlation between the total current and the current into a single absorbing patch}\label{corberg}
This cross-correlation is obtained by summing over $j$ in Eq.~(\ref{corr}) and using Eqs.~(\ref{cij}), and (\ref{barn}):
\begin{eqnarray}\nonumber
&&\quad\frac{\overline{ N_iN }-\bar{ N_i} \bar{ N}}{T}=\frac{\overline{\delta N_i\delta N }}{T} =\sum_{j=1}^sC_{ij}\\
&&=\left(\sum_{j=1}^{s}A_{ij}\right)\frac{I_2(\rho_0)}{I_1(\rho_0)},\label{ext}
\end{eqnarray} Note that the variance in (\ref{com}) can be obtained from the cross-correlation by summing over $i$ and using the definition (\ref{totalcapacitance}). We see that the dependence on the geometry in Eq.~(\ref{ext}) is factorized out as in Eq.~(\ref{com}), although via a different geometrical factor. This geometrical factor is the same as the factor which factorizes the expression for the average current into the $i$-th patch in  Eq.~(\ref{barn}).
This result is remarkable for two reasons.

First, it means that the geometry affects different gases in the same way for the purpose of calculating the average current (\ref{barn}) and the cross-correlation (\ref{ext}). For instance, changing the geometry for one gas model so as to increase the average current and the cross-correlation will have the same effect on any other gas model, in spite of their different microscopic dynamics. This property should be contrasted with the cross-correlations of currents into different patches, Eqs.~(\ref{corr}) and (\ref{mat}). There the geometry is not factorized out, so that a change in the geometry affects different gas models differently.

Second, the geometry dependence of the average current (\ref{barn}) and the cross-correlation (\ref{ext}) is factorized out via the same geometrical factor $\sum_{j=1}^{s}A_{ij}$. That is, a given gas in two different geometries but sharing the same average current into the $i$-th patch will have the same
cross-correlation between this specific current and the total current. This should again be contrasted with the cross-correlation which in general is not the same for a specified gas in two different geometries sharing the same average currents into the patches.

\subsubsection{Cell sensing by multiple receptors}\label{apberg}
Let us apply our results to the Berg-Purcell model \cite{berg} modified by an account of interactions.  In this model the cell is a  sphere of radius $a$ covered with $s\gg 1$ disk-shaped absorbing receptors with a small radius $b\ll a$, so that only a small fraction of the cell surface is covered by the disks: $sb^2/4a^2<<1$. The cell is immersed in a gas of 
diffusing molecules at density $\rho_0$. Berg and Purcell approximated the total capacitance of the system (\ref{totalcapacitance}) as $A\simeq 4\pi sab/(sb+\pi a)$. Then,
assuming that the diffusing molecules are non-interacting, they
evaluated the average total current (\ref{current2}) of molecules into the  receptors:
\begin{equation}\label{berg}
\bar{n}=\bar{n}_{\text{full}}\frac{sb}{sb+\pi a},
\end{equation}
where $\bar{n}_{\text{full}}$ is the average steady state current into a fully absorbing sphere.  Based on Eq.~(\ref{berg}), Berg and Purcell concluded the following: ``For large $s$ the intake approaches that of a completely absorbing cell, as it ought to. But it can become \emph{almost} that large before more than a small fraction of the cell's surface is occupied by absorbent patches''. Berg and Purcell gave an elegant explanation of Eq.~(\ref{berg}) in terms of a trajectory of a single diffusing molecule. Although this single-particle picture breaks down for interacting molecules, Eq.~(\ref{berg}) remains intact. Indeed, as  Eq.~(\ref{current2}) shows, the geometry is factorized out. It is also factorized out, for any gas model, in Eq.~(\ref{com}) for the variance of the total current. Therefore, if the number of absorbing patches is increased, but the area fraction of the absorbers is kept constant, fluctuations in the total current grow. Furthermore, our expression (\ref{ext})  for the cross-correlation shows how the number of patches affects cross-correlations:
\begin{eqnarray}\label{corber}
\overline{ \frac{\delta n_i\delta n}{\bar{n}_i\bar{n}} } =\frac{I_2(\rho_0)}{TI_1^3(\rho_0)}\frac{1}{A}
=\overline{\frac{\delta n^2}{\bar{n}^2} }\Big|_{\text{full}}\frac{sb+\pi a}{sb},
\end{eqnarray}
where we have used Eq.~(\ref{ext}), and $\overline{\delta n^2}/\bar{n}^2|_{\text{full}}$ is the (normalized) variance of the total current of a fully absorbing sphere. Equation~(\ref{corber}) shows that, as the number of patches increases, the cross-correlations, normalized by the mean current, decrease. This result is to be expected on physical grounds.
Importantly, it holds for any (in general, interacting) diffusive lattice gas.
\section{Optimal density and flux fields}
\label{flu2}

Here we determine the optimal profiles of the gas density and flux fields conditioned on specified absorption currents into each patch. Let us start with the gas of non-interacting RWs. Here the calculations are straightforward, see Appendix \ref{qrws}. The resulting stationary optimal density profile, $\rho_{\text{RWs}}(\mathbf{x})=\bar{\rho}_{\text{RWs}}(\mathbf{x})+\rho_{1_{\text{RWs}}}(\mathbf{x})$ can be written as
\begin{equation}
\rho_{\text{RWs}}(\mathbf{x})=\bar{\rho}_{\text{RWs}}(\mathbf{x})\left[1+\sum_{i=1}^s\frac{\delta n_i}{\bar{n}_i}\phi_i(\mathbf{x})\right],\label{qrw}
\end{equation}
where, using Eqs.~(\ref{phirw}) and (\ref{phii}), we have:
\begin{equation}
\bar{\rho}_{\text{RWs}}(\mathbf{x})=\rho_0\left[1-\sum_{i=1}^s\phi_i(\mathbf{x})\right].\label{rorw}
\end{equation}
The optimal flux field is  (see Appendix \ref{qrws}):
\begin{eqnarray}\label{jrw}
\frac{\boldsymbol{J_{\text{RWs}}}}{D_0\rho_0}
\!=\!\frac{\bar{\rho}_{\text{RWs}}}{\rho_0}\left(\sum_{i=1}^s\frac{n_i}{\bar{n}_i}\nabla\phi_i\right)
\!-\!\frac{\nabla\bar{\rho}_{\text{RWs}}}{\rho_0}\left(\sum_{i=1}^s\frac{n_i}{\bar{n}_i}\phi_i\right).
\end{eqnarray}
Let us calculate the vorticity $\nabla\times\boldsymbol{J_{\text{RWs}}}$. Taking the curl
of both sides of Eq.~(\ref{jrw}) we obtain, after cancellation of two terms,
\begin{eqnarray}
\frac{\nabla\times\boldsymbol{J_{\text{RWs}}}}{D_0\rho_0}
=2\left(\sum_{i=1}^s\frac{n_i}{\bar{n}_i}\nabla\phi_i\right)\times\left(\sum_{i=1}^s\nabla\phi_i\right).\label{vorrw}
\end{eqnarray}
This quantity is, in general, non-zero. That is, the optimal fluctuating flux field is,
in general, \emph{not} a potential vector field, in contrast to the average flux field (\ref{j}).

Equation~(\ref{vorrw}) can be generalized to an arbitrary lattice gas. For small fluctuations
the flux field is equal to
$$
\mathbf{J}=\mathbf{\bar{J}}-\nabla\left[D(\bar{\rho})\rho_1\right]+\sigma(\bar{\rho})\nabla p_1,
$$
see Eq.~(\ref{qstatglin}). The first two terms on this right-hand-side are potential vector fields, so a non-zero contribution to $\nabla \times \boldsymbol{J}$ can come only from the last term. Substituting $p_1$ from Eq.~(\ref{p1}) and using Eqs.~(\ref{phi}) and $(\ref{phii})$, we obtain
\begin{eqnarray}
\frac{\nabla\times\boldsymbol{ J}}{V(\rho_0)}
=\frac{\sigma^{\prime}(\bar{\rho})}{D(\bar{\rho})}\left(\sum_{i=1}^s\lambda_i\nabla\phi_i\right)\times\left(\sum_{i=1}^s\nabla\phi_i\right),
\label{vortex}
\end{eqnarray}
The explicit result (\ref{vorrw}) for the RWs is a particular limit of this relation. As is clear from Eq.~(\ref{vortex}), for a fluctuating system (that is, when $\lambda_i\neq 0$ for some $i$) the vorticity vanishes if and only if the two vector fields, $\sum_{i=1}^s\lambda_i \nabla\phi_i$ and $\sum_{i=1}^s\nabla\phi_i$, entering Eq.~(\ref{vortex}), are parallel. This happens in a single-current system, that is if there is only one absorbing patch. (This also happens in the extensively studied finite two-reservoir system, also sustaining a single current.) The vorticity also vanishes
if all $\lambda_i$ are equal to each other. As shown in the previous Sec.~\ref{tot}, this special case appears when the process is conditioned
on the \emph{total} absorption current into all patches combined, $n=\sum_{i=1}^s{n_i}$. We will return to this special case in Sec.~\ref{flutot}.

Let us examine the large-$|\mathbf{x}|$ asymptotic of the optimal vorticity field. The fields $\nabla\phi_i$, comprising the cross product in Eq.~(\ref{vortex}), can be expanded in multipoles, see e.g. Ref. \cite{jec}, Chap. 4. The two monopole terms, each decaying as $|\mathbf{x}|^{-2}$, do not contribute to the cross product, as they are directed in the radial direction. The leading contribution to the vorticity  comes from the cross product of a monopole field of one potential and a dipole field of the other. This product decays as $|\mathbf{x}|^{-5}$ regardless of the problem's geometry or specific lattice gas.

Bodineau \textit{et al}. \cite{partial} considered a finite two-dimensional lattice gas in  contact with two particle reservoirs. They studied fluctuations of the partial current through an imaginary slit located in the bulk and showed that the fluctuations are dominated by \emph{point-like} flux vortices localized at the edges of the slit. The action, evaluated over these solutions, can be made arbitrarily small indicating a breakdown of the MFT. In our system the flux-field vorticity is macroscopic, and is generated in the bulk, while point-like vortices are forbidden by the boundary conditions on the absorbing-reflecting surface.

\subsection{Two hemispheres}

\begin{figure}
	\includegraphics[width=0.50\textwidth,clip=]{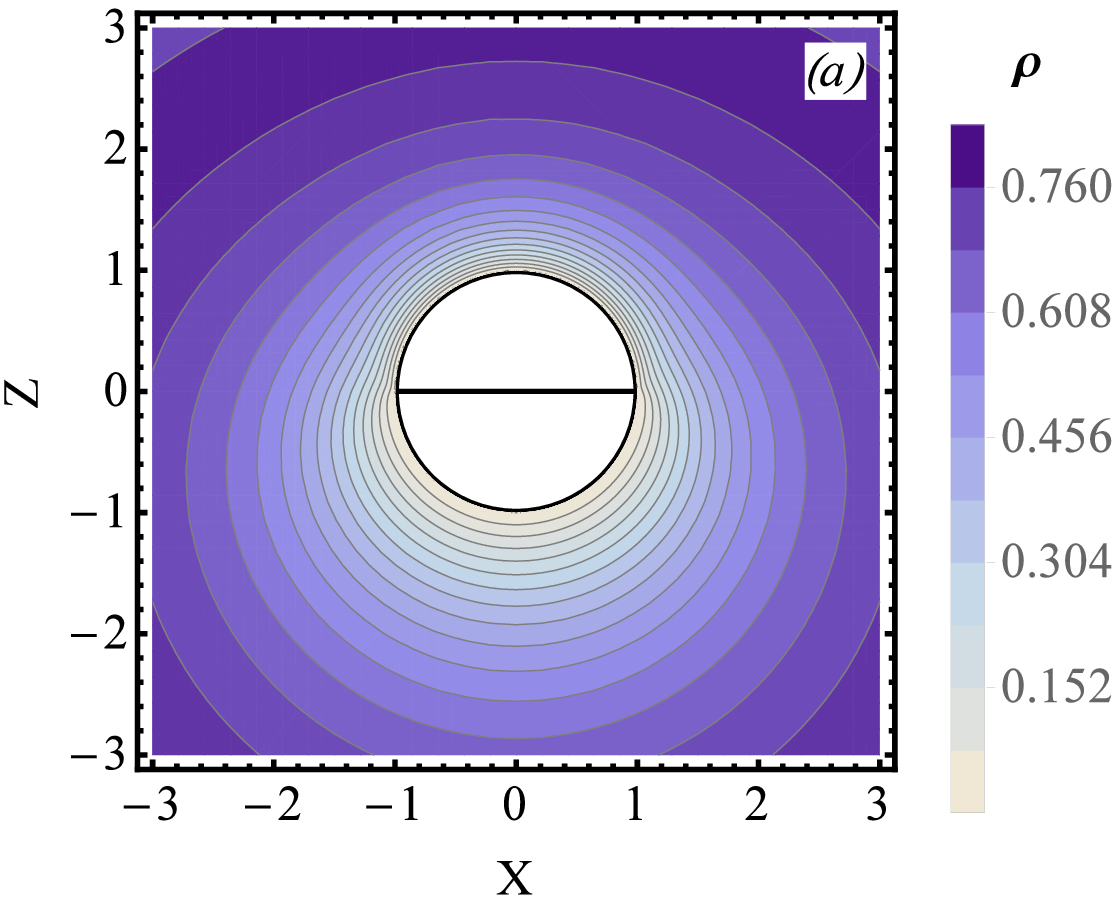}
	\includegraphics[width=0.50\textwidth,clip=]{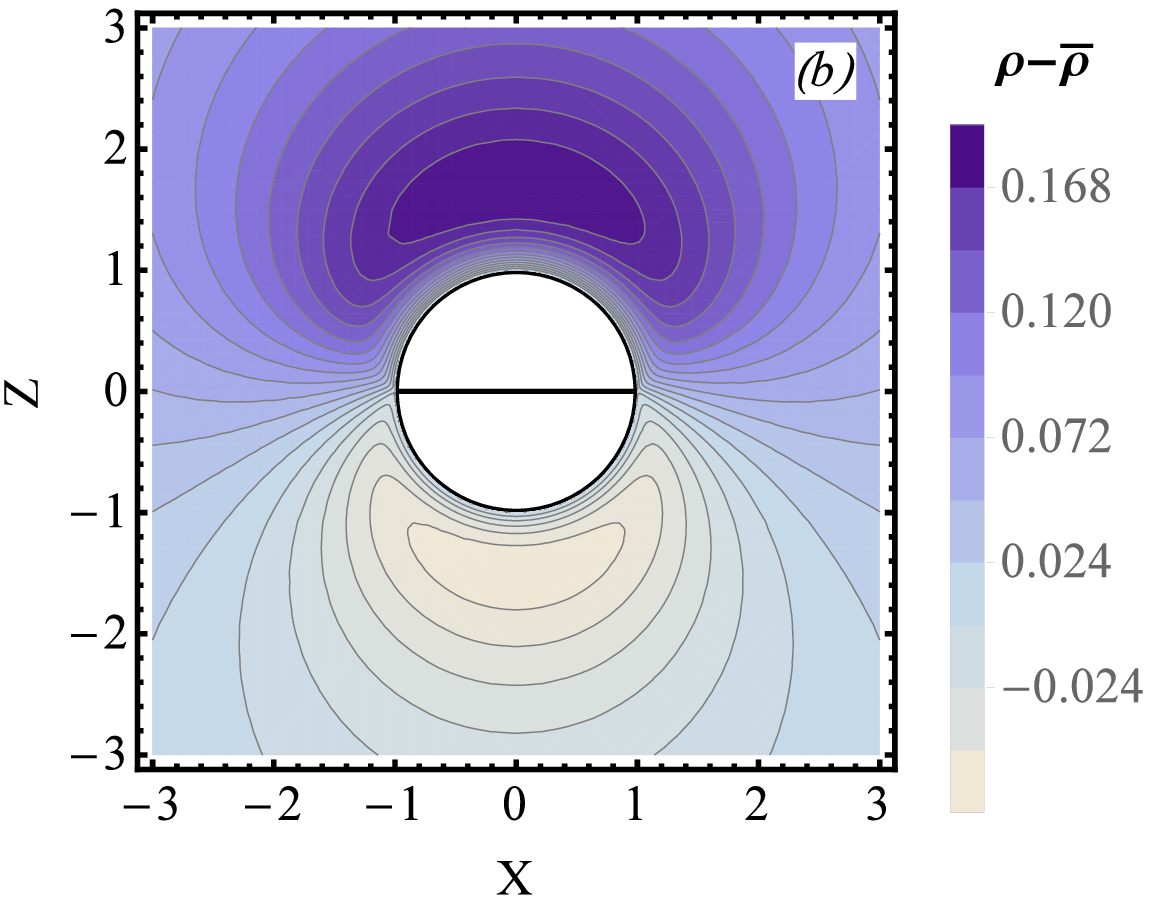}
	\caption{(a) The vertical cross section $y=0$ of the optimal density field~(\ref{qrw}) of a gas of RWs in contact with a sphere, conditioned on an enhanced particle absorption by the northern hemisphere, $n_1/\bar{n}_1=2$, and a reduced absorption by the southern hemisphere, $n_2/\bar{n}_2=0.5$, where $\bar{n}_1=\bar{n}_2=2\pi D_0\rho_0 R$. The rest of the parameters are $R=1, n_0=1$ and $D_0=1$.  The black segment is the equator. (b) The density difference $\rho_{\text{RWs}}(\mathbf{x})-\bar{\rho}_{\text{RWs}}(\mathbf{x})$. Notice the enhanced density next to the northern hemisphere and reduced density next to the southern hemisphere.}
	\label{qrwfig}
\end{figure}

\begin{figure}
	\includegraphics[width=0.50\textwidth,clip=]{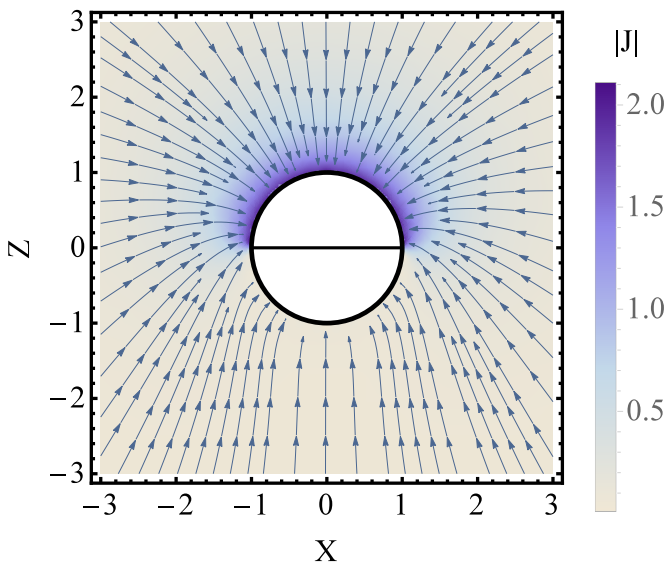}
	\caption{The flux field $\mathbf{J}_{\text{RWs}}(\mathbf{x})$ from Eq.~(\ref{jrw}) for the same setting as in Fig.~\ref{qrwfig}. 
The arrows represent the field lines of the flux.}
	\label{jrwfig}
\end{figure}

\begin{figure}
\includegraphics[width=0.50\textwidth,clip=]{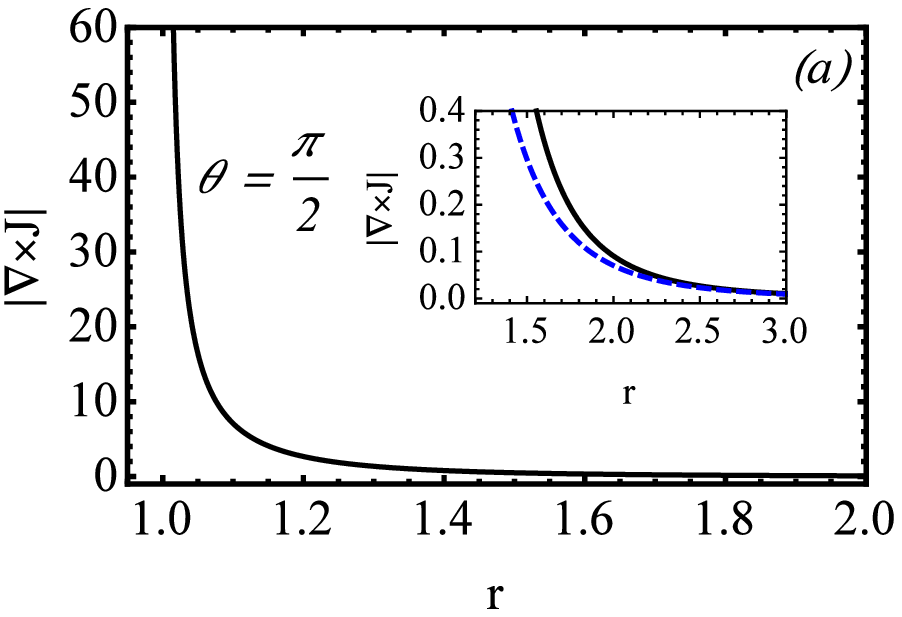}
\includegraphics[width=0.50\textwidth,clip=]{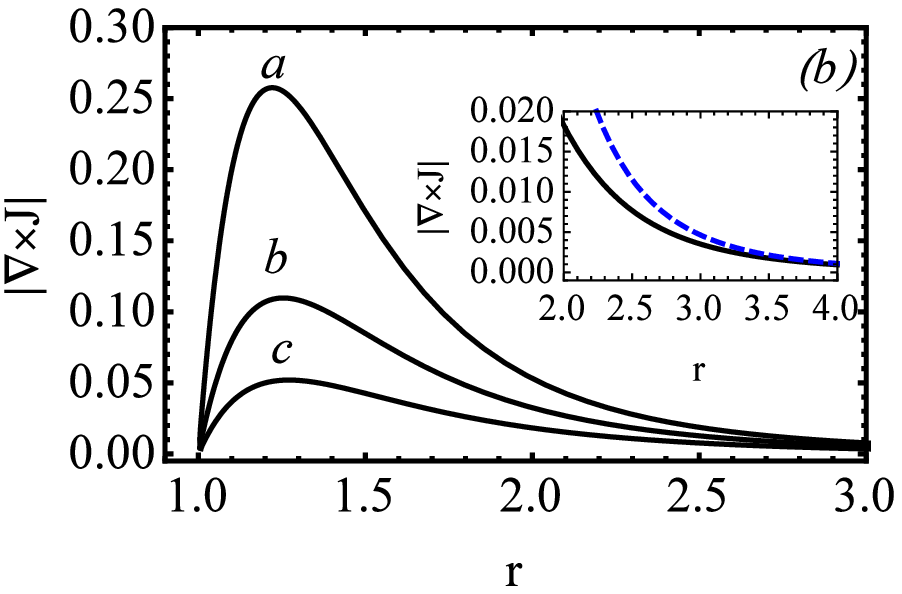}
\caption{The vorticity norm $|\nabla\times\mathbf{J}_{\text{RWs}}(r,\theta)|$, see Eq.~(\ref{vortex2}) for the same setting as in Figs.~\ref{qrwfig} and~\ref{jrwfig}.  Shown is the vorticity norm vs. $r$ in a vertical cross section for different $\theta$. (a) $\theta=\pi/2$. The vorticity diverges at $r\to R$, as predicted by Eq.~(\ref{sing}). The dashed line in the inset is the $r^{-5}$ asymptotic~(\ref{vortex3}). (b) $\theta=\pi/3$, $\pi/4$ and $\pi/6$ (marked by a, b and c,  respectively). Along these directions the vorticity vanishes on the sphere, as again predicted by Eq.~(\ref{sing}). The dashed line in the inset is the $r^{-5}$ asymptotic~(\ref{vortex3})  for $\theta=\pi/6$.}
\label{rotjrwfig}
\end{figure}

To illustrate our results on the optimal density and flux fields, we consider a gas of RWs in contact with a sphere of radius $R$ placed in the origin. The whole sphere is absorbing,
and we are interested in the absorption  statistics of each of the two hemispheres: the northern one, $\theta\in\left[0,\pi/2\right]$, which we denote by $\Omega_1$ and the southern one, $\theta\in\left[\pi/2, \pi\right]$, denoted by $\Omega_2$. Here $\theta$ is the polar angle of the spherical coordinates.
The problem possesses cylindrical symmetry. The effective potential, defined in Eq.~(\ref{phi}), is $\phi=D_0\rho_0R/r$, where $r$ is the radial coordinate. The average absorption currents (\ref{current}) to each of the two hemispheres are identical and equal to $\bar{n}_1=\bar{n}_2=2\pi D_0\rho_0 R$.
The set of effective potentials  $\phi_i$
can be found explicitly via expansion in spherical harmonics:
\begin{equation}\label{hemi}
\phi_{1,2}=\frac{R}{2r}\pm f(r,\theta),
\end{equation}
where the plus and minus signs refer to the northern and southern hemispheres, respectively. The function $f(r,\theta)$ is given by
\begin{equation}\label{hemif}
f(r,\theta)\!=\!\frac{\sqrt{\pi}}{4}\sum_{k=1}^{\infty}\frac{\left(4k-1\right)}{\Gamma\left(\frac{3}{2}-k\right)\Gamma(1+k)} \left(\frac{R}{r}\right)^{2k} \!\!P_{2k-1} (\cos \theta),
\end{equation}
$P_{2k-1}(\dots)$ are the Legendre polynomials, and $\Gamma(\dots)$ is the gamma function. Plugging expressions (\ref{hemi}) and (\ref{hemif}) in Eq.~(\ref{vorrw}) we obtain,
after some algebra, the vorticity:
\begin{equation}
\nabla\times\boldsymbol{ J}_{\text{RWs}}=\frac{n_1-n_2}{\pi r^3}\frac{\partial f}{\partial\theta}\hat{\varphi}\label{vortex2}
\end{equation}
where $\hat{\varphi}$ is the unit vector in the azimuthal direction $\varphi$ in spherical coordinates.  Using Eq.~(\ref{hemif}), we can obtain the large-$r$ behavior of the vorticity:
\begin{equation}
\label{vortex3}
\nabla\times\boldsymbol{J}_{\text{RWs}}(r\to \infty) \simeq -\frac{3R^2 \left(n_1-n_2\right) \sin \theta}{4\pi r^5} \hat{\varphi} .
\end{equation}
The $r^{-5}$ dependence is expected from the general argument given above. The vorticity exhibits a delta-function singularity at the equator of the sphere:
\begin{equation}
\nabla\times\boldsymbol{J}_{\text{RWs}}(r\to R) = -\frac{n_1-n_2}{\pi R^3}\,\delta\left(\theta-\frac{\pi}{2}\right)\hat{\varphi}.
\label{sing}
\end{equation}
We can also find the optimal flux and optimal density fields in this example by plugging expressions (\ref{hemi}) and (\ref{hemif}) in Eqs.~(\ref{qrw})-(\ref{jrw}). Figure~\ref{qrwfig} shows a vertical cross-section of the optimal density field $\rho_{\text{RWs}}$ for $n_1/\bar{n}_1=2$ and $n_2/\bar{n}_2=0.5$. As intuitively expected, the optimal density is higher/lower than the mean density close to the northern/southern hemisphere, respectively. The optimal flux field~(\ref{jrw}) and the vorticity field (\ref{vortex2}) in this case are shown in Figs.~\ref{jrwfig} and~\ref{rotjrwfig}, respectively.  Evident in  Fig. \ref{rotjrwfig} is the singularity of the vorticity at $r\rightarrow R$, displayed by Eq.~(\ref{sing}). The flux field (\ref{jrw}) exhibits a milder singularity at $r\rightarrow R$, where the radial component of the flux is discontinuous as a function of $\theta$ at $\theta=\pi/2$. These singularities are due to the discontinuous boundary conditions for the potentials $\phi_i$'s, and they are generic when considering several patches (or reservoirs) in direct contact with each other.  The density field (\ref{qrw}) is continuous at $r=R$.

\subsection{Optimal profiles conditioned on the total absorption current}\label{flutot}
The optimal density and flux fields, conditioned on the total absorption current deviation $\delta n$,
are given by the solution of the linearized MFT equations (\ref{qstatglin}) and (\ref{pstatglin}) alongside with Eqs.~(\ref{optn}) and~(\ref{optlam}). In this case the equations can be explicitly solved for any lattice gas, see Appendix \ref{totqp}, and we obtain
\begin{eqnarray}
\!\!\!\!\!\!\!\!\!\!D(\bar{\rho})\rho_1&=&\frac{\delta n}{\bar{n}}\left[\frac{\tilde{I}_1(\bar{\rho},\rho_0)I_2(\rho_0)-I_1(\rho_0)\tilde{I}_2(\bar{\rho},\rho_0) }{I_2(\rho_0)}\right], \label{qq}\\
\!\!\!\!\!\!\!\!\!\!p_1&=&\frac{\delta n}{\bar{n}}\frac{I_1(\rho_0)}{I_2(\rho_0)}\tilde{I}_1(\bar{\rho},\rho_0),\label{pp}
\end{eqnarray}
where
\begin{eqnarray}\label{itild}
\tilde{I}_k\left[\bar{\rho}(\mathbf{x}),\rho_0\right]\equiv\int_{\bar{\rho}(\mathbf{x})}^{\rho_0}D(w)\sigma(w)^{k-1}dw.
\end{eqnarray}
Note that $\phi\left(\mathbf{x}\right)=\tilde{I}_1\left[\bar{\rho}\left(\mathbf{x}\right),\rho_0\right]$, see Eq.~(\ref{phi}).
As Eqs.~(\ref{qq}) and (\ref{pp}) show, both $\rho(\mathbf{x})=\bar{\rho}(\mathbf{x})+\rho_1(\mathbf{x})$, and $p(\mathbf{x})=p_1(\mathbf{x})$ depend on $\bar{\rho}(\mathbf{x})$ alone. Remarkably, these $\bar{\rho}(\mathbf{x})$-dependencies are described by (the small-fluctuation limit of) the optimal profiles $\rho^{(1)}(x)$ and $p^{(1)}(x)$ of the \emph{one-dimensional} problem  for the reservoir density values $\rho_a=\rho_0$ and $\rho_b=0$, found in
Ref. \cite{bd}:
\begin{eqnarray}\nonumber\label{q1d}
&&\rho(\mathbf{x})=\rho^{(1)}\left[\phi\left[\bar{\rho}\left(\mathbf{x}\right)\right]\right],\\ &&p(\mathbf{x})=p^{(1)}\left[\phi\left[\bar{\rho}\left(\mathbf{x}\right)\right]\right],
\end{eqnarray}
where $\phi\left(\bar{\rho}\right)$ is defined in Eq.~(\ref{phi}). As one can see from Eq.~(\ref{q1d}), the average density $\bar{\rho}(\mathbf{x})$ plays the role of the natural ``spatial coordinate" of the problem. (The same property
was uncovered and exploited in a different geometry in  Ref. \cite{shpiel}.) With respect to this spatial coordinate the problem is effectively one-dimensional, hence the relation to the current fluctuations of the one-dimensional system, described at the end of Sec.  \ref{tot}.

Finally, using Eqs.~(\ref{qq}) and~(\ref{pp}), we obtain the optimal fluctuating flux field (\ref{qstatglin}), conditioned on
the total absorption current:
\begin{equation}\label{totflux}
\mathbf{J}=\frac{n}{\bar{n}}\nabla\tilde{I}_1(\bar{\rho},\rho_0)=\frac{n}{\bar{n}}\bar{\mathbf{J}},
\end{equation}
where $\bar{\mathbf{J}}$ is the average steady state flux field (\ref{j}). As Eq.~(\ref{totflux}) shows, in order to generate a total absorption current which is larger by a factor $n/\bar{n}$ than the average one, the system has to increase the optimal flux by the same factor everywhere in space (see also Refs. \cite{fullabsorb,shpiel}). It is not surprising, therefore, that the optimal flux field, conditioned on the total absorption current, is vortex-free.

\section{Discussion}\label{conc}

In this work we employed the Macroscopic Fluctuation Theory (MFT) to determine the statistics of particle absorption by several patches located on the surface of a domain immersed in a gas composed of interacting diffusing particles. Essentially, this work extends a number of previous results \cite{Derrida2007,MFTreview,bd,fullabsorb,shpiel}, obtained for different single-current settings, to a multiple-current setting. Among central results of this work are Eqs.~(\ref{mat}) and (\ref{offdiag}) for the covariance matrix of the absorption currents into different patches, and Eq.~(\ref{condition}) which establishes, independently of the system's geometry, whether the absorption currents correlate or anti-correlate.
The same condition (\ref{condition})  has recently gained much attention in a different context -- as a sufficient condition for the validity of the additivity hypothesis for arbitrary currents \cite{main,ber,main2}. This coincidence hints at a possible relation between correlations and dynamical phase transitions  via which the additivity property breaks down.

One particular example of diffusive transport, extensively studied in the past,
describes electronic transport in mesoscopic wires \cite{meso}. In this case Eq.~(\ref{lang}), with $D$ and $\sigma$ corresponding to the SSEP, serves as a suitable mathematical description \cite{jor}. As follows from Eq.~(\ref{condition}), the cross-correlations in this system are strictly negative. This result has been known for some time: it was obtained both by the scattering matrix method \cite{but} and by the effective Langevin description \cite{multi}. It has been verified in experiment and gained much attention, see \textit{e.g.} \cite{f}. It is often referred to as the electronic Hanbury Brown and Twiss effect \cite{meso}, and is intimately related to the Fermi statistics of charge carriers.
The MFT framework which we developed here enables one to address a much broader class of diffusive systems by considering more general diffusivity and mobility. As we have seen, this may lead to qualitatively different fluctuational behaviors, such as different signs of the correlation terms.

An important example of positive correlations in the (energy) absorption appears in the context of diffusive wave propagation in disordered media \cite{penini,main}. Indeed, mesoscopic wave transport can often be described by Eq.~(\ref{lang}) which relates the wave energy flux to the local wave intensity $I$: an equivalent of particle density $\rho$. Remarkably, this ``lattice gas" can be described in terms of the KMP model. Indeed, the corresponding diffusion coefficient is constant here, but the mobility  behaves as  $\sigma(I)\propto I^2$  \cite{penini}. In this case our Eq.~(\ref{condition}) predicts strictly positive cross-correlations in the photon absorption.
It would be very interesting to test this prediction in experiment. The experimental setting can be similar to that of Ref. \cite{shapiro2},  where a waveguide filled with a disordered material was used to study the light-intensity correlations between two distant points along the waveguide.

Another important advantage of the MFT is that it predicts, for all settings,  the optimal (most probable) spatial profiles of the density and of the flux, conditioned on a given multiple-current statistics. As we have shown here, the corresponding optimal flux field generically exhibits vorticity. The vorticity disappears if  the process is conditioned on the \emph{total} absorption current into all the patches combined. It would be interesting to see whether the vorticity still disappears for the total absorption current if one goes beyond typical, small fluctuations which we addressed here. We found that, \emph{under the assumption} that the optimal flux field is vortex-free, the simple solution (\ref{q1d}) (which solves the time-independent \emph{nonlinear} MFT equations exactly), is unique. As a result all the cumulants [and not only the first two as in  Eq.~(\ref{com})],  are equal to the corresponding one-dimensional cumulants times the capacitance of the system. The irrotational character of the flux field is, however, a strong assumption. Its breaking, at a critical value of the total absorption current, may have a character of phase transition.

The MFT formalism can be readily extended to other geometries sustaining multiple currents in systems of interacting diffusing particles. It  can also be extended to \emph{finite} two-dimensional systems where a set of characteristic potentials $\phi_i$ and the capacitance matrix can always be defined. An \emph{infinite} two-dimensional system does not reach a true steady state, and logarithmic corrections to the linear scaling of the action with time $T$ are to be expected \cite{b}.  Last but not least, the formalism can also be extended, with some modifications, to the case when different reservoirs are kept at different densities.

\section*{ACKNOWLEDGMENTS}
We are grateful to  Eric Akkermans, Yaron Bromberg, Giovanni Jona-Lasinio and  Naftali Smith for useful discussions of different parts of this work.
We acknowledge financial support from the Israel Science Foundation (grant No. 807/16) and the United States-Israel Binational Science Foundation
(BSF) (grant No. 2012145).

\appendix

\section{Derivation of the MFT equations and boundary conditions}\label{mform}		

There are several methods for derivation of the MFT equations \cite{MFTreview}. Here we use the Martin-Siggia-Rose formalism \cite{MFTreview,MSR,deridamft,map}. We start from Eq.~(\ref{lang}) and represent the probability of observing a joint density and flux history $\rho(\mathbf{x},t), \mathbf{J}(\mathbf{x},t)$, constrained by the conservation law $\partial_{t}{\rho}+\nabla \cdot \mathbf{J}=0$,
as a path integral:
\begin{eqnarray}\label{path1}
&&\mathcal P \simeq\int\mathcal{D}\rho\mathcal{D}\mathbf{J}\prod_{\mathbf{x},t}\delta(\partial_{t}{\rho}+\nabla \cdot \mathbf{J})\\
&&\times \exp\left\{-\int_0^Tdt\int d\mathbf{x}\frac{\left[\mathbf{J}+D(\rho)\nabla \rho\right]^2}{2\sigma(\rho)}\right\}.\nonumber
\end{eqnarray}
Using an integral representation for the $\delta$-function with the help of an auxiliary field $p(\mathbf{x},t)$, we can rewrite Eq.~(\ref{path1}) as a path integral over three unconstrained fields \cite{map}:
\begin{eqnarray}\label{act}\nonumber
&&\mathcal P \simeq\int\mathcal{D}\rho\mathcal{D}\mathbf{J}\mathcal{D}p\exp\left\{-\mathcal{ L}\left[\rho(\mathbf{x},t),\mathbf{J}(\mathbf{x},t),p(\mathbf{x},t)\right] \right\}, \\\nonumber
&&\mathcal {L}=\int_0^Tdt\int d\mathbf{x}\left\{\frac{\left[\mathbf{J}+D(\rho)\nabla \rho\right]^2}{2\sigma(\rho)}+p\left(\partial_{t}{\rho}+\nabla \cdot \mathbf{J}\right)\right\}.\nonumber\\
\end{eqnarray}
We wish to evaluate the path integral over only those histories which led to $N_i$ particles being absorbed, by time $T$, by the $i$-th patch, $i=1,2, \dots, s$.
Assuming that all characteristic length scales are macroscopic and include large numbers of particles, we can evaluate this path integral via a saddle-point approximation. The dominant contribution comes from the optimal fluctuation: the most probable history $(\rho,\mathbf{J},p)$
leading to a specified number of absorbed particles $N_i$. The problem, therefore, reduces to finding the minimum
of $\mathcal L$ under the $s$ constraints:
\begin{eqnarray}\label{cons}
N_i=\int_0^Tdt\oint_{\Omega_i}\mathbf{J}\cdot \hat{n}ds ,\quad i=1,2, \dots, s.
\end{eqnarray}
The latter can be incorporated via $s$ Lagrange multipliers $\lambda_i$, $i=1,2, \dots, s$:
\begin{multline}\label{mform2}
-\ln{\mathcal P}(N_1,N_2,\dots,N_s;\rho_0,T) \simeq S,\\
S=\min_{\rho,\mathbf{J},p}\biggl\{\mathcal{ L}\left[\rho(\mathbf{x},t),\mathbf{J}(\mathbf{x},t),p(\mathbf{x},t)\right]\\
+\sum_i\lambda_iN_i-\sum_i\lambda_i\int_0^Tdt\oint_{\Omega_i}\mathbf{J}\cdot \hat{n}ds\biggr\},
\end{multline}
where the \textit{a priori} unknown Lagrange multipliers are ultimately set by the $s$ constraints (\ref{cons}).
Taking the first variation of $S$ we obtain:
\begin{multline}\label{deltas}
\delta S=\int_0^Tdt\int d\mathbf{x}\delta \rho\biggl\{-\partial_{t}{p}-\frac{1}{2}\sigma^{\prime}\left(\rho\right)\left[\frac{\mathbf{J}+D(\rho)\nabla \rho}{\sigma(\rho)}\right]^2\\
-D(\rho)\nabla\cdot\left[\frac{\mathbf{J}+D(\rho)\nabla \rho}{\sigma(\rho)}\right]\biggr\}\\
+\int_0^Tdt\oint_{\Omega_r}\delta \rho \frac{D^2(\rho)}{\sigma(\rho)}\nabla \rho\cdot\hat{n} ds+\int d\mathbf{x}\delta \rho\, p|_{t=T}\\
+\int_0^Tdt\int d\mathbf{x}\boldsymbol{\delta J}\cdot\left[\frac{\mathbf{J}+D(\rho)\nabla \rho-\sigma(\rho)\nabla p}{\sigma(\rho)}\right]\\
+\sum_i\int_0^Tdt\oint_{\Omega_i}\left(p-\lambda_i\right)\boldsymbol{\delta J}\cdot\hat{n} ds,\\
+\int_0^Tdt\int d\mathbf{x}\delta p\left(\partial_{t}{\rho}-\nabla\cdot \mathbf{J}\right).
\end{multline}
Here we have used the conditions
\begin{eqnarray}\nonumber\label{bound}
&&\delta \rho(\mathbf{x},t=0)=\delta \rho(\mathbf{x}\in\Omega_i,t)=\delta \rho(\mathbf{x}\rightarrow\infty,t)\\
&&=\boldsymbol{\delta J}(\mathbf{x}\in{\Omega_r},t)\cdot\hat{n}=0,
\end{eqnarray}
following from the boundary conditions (\ref{blang})-(\ref{blang3}).  We have also used Gauss's theorem to transform volume integrals to surface integrals. There are no contributions from the surface integral at infinity.

Setting $\delta S$ to zero and using the fact that $\delta \rho(\mathbf{x}\in\Omega_r,t)$, $\delta \rho (\mathbf{x},t=T)$, and $\boldsymbol{\delta J}(\mathbf{x}\in\Omega_i,t)$
can be arbitrary, we see that the the optimal flux field is given by
\begin{equation}\label{jj}
\mathbf{J}=-D(\rho) \nabla \rho+\sigma(\rho) \nabla p,
\end{equation}
and the optimal profiles $\rho$ and $p$ satisfy the MFT equations (\ref{d11}) and (\ref{d12}) in the bulk and the boundary conditions (\ref{bt0})-(\ref{bcgen13}).
The MFT equations (\ref{d11}) and (\ref{d12}) are Hamiltonian, where $\rho$ and $p$ play the role of the conjugate ``coordinate" and ``momentum" density fields.
The Hamiltonian is
\begin{eqnarray}\label{hamilton1}
H\left[\rho(\mathbf{x},t),p(\mathbf{x},t)\right]=\int d\mathbf{x}\,\mathcal{H},
\end{eqnarray}
where the Hamiltonian density is given by \cite{MFTreview}
\begin{eqnarray}\label{hamilton2}
\mathcal{H}=\frac{\sigma(\rho)\left(\nabla p\right)^2}{2}-D(\rho)\nabla \rho\cdot\nabla p.
\end{eqnarray}
Finally, the action (\ref{mform2}) can be simplified to~(\ref{actionmain}) by using Eqs.~(\ref{d11}),~(\ref{cons}) and~(\ref{jj}).

\section{Derivation of Eq.~(\ref{sigmalin})}\label{lamcur}

The formal solution to Eq.~(\ref{pois}) is given by Green's function $G(\mathbf{x},\mathbf{x'})$ which satisfies the equations
\begin{eqnarray}
\nabla^2G&=&\delta(\mathbf{x}-\mathbf{x'}),\label{delta}\\
G(\mathbf{x}\in\Omega_i,\mathbf{x'})&=&0,\label{omi}\\
\nabla_{\mathbf{x}}G(\mathbf{x}\in\Omega_r,\mathbf{x'})\cdot\hat{n}&=&0,\label{omr}\\
G(\mathbf{x}\rightarrow\infty,\mathbf{x'})&=&0.
\end{eqnarray}
$G(\mathbf{x},\mathbf{x'})$  can be interpreted as the electrostatic potential at point $\mathbf{x}$, induced by a point charge $q=-1/4\pi$ at point $\mathbf{x'}$ outside the domain, when all the patches $\Omega_i$ are conducting and grounded. With  $G(\mathbf{x},\mathbf{x'})$,  the solution to Eq.~(\ref{pois}) can be written as
\begin{equation}
\left[D(\bar{\rho})\rho_1\right](\mathbf{x})=\int d\mathbf{x'}G(\mathbf{x},\mathbf{x'})\nabla\cdot\left[\sigma(\bar{\rho})\nabla p_1\right](\mathbf{x'}).\label{qg}
\end{equation}
Now let us evaluate the flux deviation
\begin{equation}\nonumber
\boldsymbol{\delta J}=-\nabla\left[D(\bar{\rho})\rho_1\right]+\sigma(\bar{\rho})\nabla p_1,
\end{equation}
whose surface integrals over the patches give the corresponding current deviations:
\begin{eqnarray}\nonumber
&&\delta n_i=\oint_{\Omega_i} \left\{-\nabla\left[D(\bar{\rho})\rho_1\right]+\sigma(\bar{\rho})\nabla p_1\right\}\cdot\hat{n}dS=\\\nonumber
&&-\int d\mathbf{x'}\nabla\cdot\left[\sigma(\bar{\rho})\nabla p_1\right](\mathbf{x'})\oint_{\Omega_i}\!\!\!\!\!\nabla_{\mathbf{x}}G(\mathbf{x},\mathbf{x'})\cdot\hat{n}dS\\
&&+\oint_{\Omega_i}\sigma(\bar{\rho})\nabla p_1\cdot\hat{n}dS.
\end{eqnarray}
Using a Green's function identity
\begin{equation}
\oint_{\Omega_i}\nabla_{\mathbf{x}}G(\mathbf{x},\mathbf{x'})\cdot\hat{n}dS=\phi_i(\mathbf{x'})
\end{equation}
 \cite{jec}, we obtain
\begin{eqnarray}\nonumber\label{delta1}
\delta n_i=
-\int d\mathbf{x}\nabla \cdot \left[\sigma(\bar{\rho})\nabla p_1\right]\phi_i +\oint_{\Omega_i}\sigma(\bar{\rho})\nabla p_1\cdot\hat{n}dS.\\
\end{eqnarray}
As the last step, we use Gauss's theorem to transform the surface integral in Eq.~(\ref{delta1}) to a volume integral with the help of the potential $\phi_i$
defined in Sec.~\ref{m}:
\begin{eqnarray}\nonumber\label{delta2}
&&\delta n_i=
\int d\mathbf{x} \Big\{-\nabla \cdot \left[\sigma(\bar{\rho})\nabla p_1\right]\phi_i +\nabla \cdot \left[\sigma(\bar{\rho})\nabla p_1\phi_i\right]\Big\}\\\nonumber
&&=\int d\mathbf{x}\,\sigma(\bar{\rho})\nabla p_1\cdot\nabla\phi_i=\sum_j\lambda_j\int d\mathbf{x}\,\sigma(\bar{\rho})\nabla\phi_j\cdot\nabla\phi_i,\\
\end{eqnarray}
where in the last equality we have substituted $p_1$ from Eq.~(\ref{p1}). This linear relation can be written in a matrix form as in Eq.~(\ref{sigmalin}).

\section{Matrix $\boldsymbol C$ is positive definite}\label{mat1}
To prove this statement one needs to show that, for every nontrivial  $s$-dimensional vector $\boldsymbol V$,
$\boldsymbol V^T\cdot \boldsymbol C \cdot \boldsymbol V> 0$. We have
\begin{eqnarray}\nonumber
&&\boldsymbol V^T\cdot \boldsymbol C \cdot \boldsymbol V=\sum_{i,j}\int d\mathbf{x}\,
V_iV_j\sigma\left(\bar{\rho}\right)\nabla\phi_i\cdot\nabla\phi_j\\
&&=\int d\mathbf{x}\,
\sigma\left(\bar{\rho}\right)\left[\sum_{i}V_i\nabla\phi_i\right]^2,\label{bform}
\end{eqnarray}
which is clearly non-negative. Furthermore, since the fields $\nabla\phi_i$ are linearly independent, it is positive definite.

\section{Derivation of Eq.~(\ref{offdiag})}\label{offdiag1}
We start with rewriting (\ref{mat}):
\begin{eqnarray}\nonumber\label{2}
&&\int d\mathbf{x}\,\sigma(\bar{\rho})\nabla\phi_i\cdot\nabla\phi_j=\int d\mathbf{x}\,\frac{1}{2}\phi_i\phi_j\nabla^2 \sigma(\bar{\rho})\\
&&+\int d\mathbf{x}\,\nabla \cdot \left[\frac{\sigma(\bar{\rho})}{2}\nabla\left(\phi_i\phi_j\right)-\phi_i\phi_j\frac{\nabla\sigma(\bar{\rho})}{2}\right],
\end{eqnarray}
where we have used the fact that $\phi_i$-s are harmonic functions. Using Gauss's theorem, we can transform the last integral
into a surface integral. There is no contribution from the surface integral at infinity since both $\nabla\left(\phi_i\phi_j\right)$, and $\phi_i\phi_j\nabla\sigma(\bar{\rho})$ decay as $\mathbf{x}^{-3}$ as $\mathbf{x}\rightarrow\infty$. Therefore, the last integral in Eq.~(\ref{2}) becomes
\begin{eqnarray}\nonumber
&&\int d\mathbf{x}\,\nabla \cdot \left[\frac{\sigma(\bar{\rho})}{2}\nabla\left(\phi_i\phi_j\right)-\phi_i\phi_j\frac{\nabla\sigma(\bar{\rho})}{2}\right]\\
&&=\oint_{\Omega} \left[\frac{\sigma(\bar{\rho})}{2}\nabla\left(\phi_i\phi_j\right)-\phi_i\phi_j\frac{\nabla\sigma(\bar{\rho})}{2}\right]\cdot\hat{n}dS
\label{BB}.
\end{eqnarray}
As
\begin{equation}\nonumber
\nabla\phi_i\left(\mathbf{x}\in\Omega_r\right)\cdot\hat{n}=\nabla\bar{\rho}\left(\mathbf{x}\in\Omega_r\right)\cdot\hat{n}=0,
\end{equation}
the surface integral is taken over the absorbing patches $\Omega_i$ alone. Moreover, since
\begin{equation}\nonumber
\sigma\left[\bar{\rho}\left(\mathbf{x}\in\Omega_i\right)\right]=\sigma(0)=0,
\end{equation} the surface integral of the first term vanishes. Finally, since $\phi_i\left(\mathbf{x}\in\Omega_j\right)=\delta_{i,j}$ we obtain
\begin{eqnarray}\nonumber
&&\int d\mathbf{x}\,\nabla \cdot \left[\frac{\sigma(\bar{\rho})}{2}\nabla\left(\phi_i\phi_j\right)-\phi_i\phi_j\frac{\nabla\sigma(\bar{\rho})}{2}\right]\\
&&=-\delta_{i,j}\oint_{\Omega_i}\frac{\nabla\sigma(\bar{\rho})}{2} \cdot\hat{n}dS\\\nonumber &&= \delta_{i,j}\oint_{\Omega_i}\frac{\sigma^{\prime}(\bar{\rho})}{2D(\bar{\rho})}\left[-D(\bar{\rho})\nabla\bar{\rho} \right]\cdot\hat{n}dS=\delta_{i,j}\frac{\sigma^{\prime}(0)}{2D(0)}\bar{n}_i,
\end{eqnarray}
where in the last equality we have used the fact that, on the absorbing patches, $\bar{\rho}=const=0$, and identified the average currents (\ref{current}) expressed through the average flux field (\ref{j}). What is left to arrive at Eq.~(\ref{offdiag}) is to rewrite the integrand of the second integral in Eq.~(\ref{2}) as
\begin{eqnarray}\nonumber
\frac{1}{2}\phi_i\phi_j\nabla^2\left[\sigma(\bar{\rho})\right]=\frac{1}{2}\phi_i\phi_j\frac{\left(\nabla\phi\right)^2}{D(\bar{\rho})}\left[\frac{\sigma^{\prime}(\bar{\rho})}{D(\bar{\rho})}\right]^{\prime},
\end{eqnarray}
where we have used Eq.~(\ref{phi}) and the fact that $\phi$ is a harmonic function.

\section{Conditioning on the total absorption current} \label{optimum}

We seek the minimum of the bilinear form
$$
\boldsymbol{\delta n}^T\cdot\boldsymbol C^{-1}\cdot\boldsymbol {\delta n}
$$
subject to the constraint
$$
\mathbf{\boldsymbol\delta n}^T \cdot (1,1,\dots,1)=\delta n.
$$
Introducing an additional Lagrange multiplier $\gamma$, we minimize the function
$$
\mathbf{\boldsymbol\delta n}^T\cdot \boldsymbol C^{-1} \cdot \mathbf{\boldsymbol\delta n}-\gamma\left[\mathbf{\boldsymbol\delta n}^T \cdot (1,1,\dots,1)-\delta n\right]
$$
with respect to all $\delta n_i$. The minimization yields
$$
\delta n_i=\frac{\gamma}{2}\sum_j C_{ij}.
$$
Comparing this expression with Eq.~(\ref{sigmalin}), we see that
\begin{equation}\nonumber
\lambda_i=\lambda \equiv\frac{\gamma}{2}.
\end{equation}
Now we can find the currents and all $\lambda_i$-s in terms of the total absorption current $n$.  We plug the identity
$\sum_j\phi_j=\phi/V(\rho_0)$  in Eq.~(\ref{mat}) and obtain
\begin{equation}\label{sum}
\sum_{j}C_{ij}=\frac{1}{V(\rho_0)}\int d\mathbf{x}\,
\sigma\left(\bar{\rho}\right)\nabla\phi_i\cdot\nabla\phi .
\end{equation}
Now we introduce a single-variable function  $\bar{\rho}(\phi)$ as the inverse function to $\phi(\bar{\rho})$ defined in Eq.~(\ref{phi}). Using  $\bar{\rho}(\phi)$,
we can rewrite the integrand as
\begin{equation}\nonumber
\sigma\left[\bar{\rho}(\phi)\right]\nabla\phi\cdot\nabla\phi_i=\nabla\cdot\left\{\nabla\phi_i\int_0^{\phi}\sigma\left[\bar{\rho}(w)\right]dw\right\},
\end{equation}
where we have also used the fact that $\phi_i$ is a harmonic function. Applying Gauss's theorem for this form of the integrand in Eq.~(\ref{sum}) we have (the surface integral at infinity vanishes due to our choice of the lower bound of the integral on $\sigma$, and that $\nabla\phi_i$ decays as $\mathbf{x}^{-2}$):
\begin{eqnarray}
&&\sum_{j}C_{ij}=\frac{1}{V(\rho_0)}\oint_{\Omega} \left\{\int_0^{\phi}\sigma\left[\bar{\rho}(w)\right]dw\right\}\nabla\phi_i \cdot \hat{n}dS\nonumber \\\nonumber
&&=\frac{1}{V(\rho_0)}\sum_j\oint_{\Omega_j} \left\{\int_0^{\phi}\sigma\left[\bar{\rho}(w)\right]dw\right\}\nabla\phi_i \cdot \hat{n}dS\\\nonumber
&&=\frac{\int_0^{V(\rho_0)}\sigma\left[\bar{\rho}(w)\right]dw}{V(\rho_0)}\sum_j\oint_{\Omega_j} \nabla \phi_i\cdot\hat{n}dS\\\nonumber
&&=\frac{\int_0^{V(\rho_0)}\sigma\left[\bar{\rho}(w)\right]dw}{V^2(\rho_0)}\oint_{\Omega_i} V(\rho_0)\sum_j\nabla \phi_j\cdot\hat{n}dS\\
&&=\frac{\int_0^{V(\rho_0)}\sigma\left[\bar{\rho}(w)\right]dw}{V^2(\rho_0)}\bar{n}_i= \frac{\int_0^{\rho_0}D(w)\sigma(w)dw}{V^2(\rho_0)}\bar{n}_i, \label{sumsigma1}
 \end{eqnarray}
where in the second and third equalities we used the boundary conditions:
\begin{equation}\nonumber
\nabla \phi_i(\mathbf{x}\in\Omega_r)\cdot\hat{n}=0\quad,\quad\phi(\mathbf{x}\in\Omega_j)=\text{const}=V(\rho_0),
\end{equation}
respectively. In the forth equality we employed the identity
\begin{equation}\nonumber
\oint_{\Omega_j} \nabla \phi_i\cdot\hat{n}dS=\oint_{\Omega_i} \nabla \phi_j\cdot\hat{n}dS,
\end{equation}
which reflects the symmetry of the capacitance matrix (\ref{a}).
As a result,
\begin{equation}\label{111}
\delta n_i=\frac{\gamma}{2}\alpha(\rho_0) \bar{n}_i,
\end{equation}
where $\alpha(\rho_0)$ is defined in (\ref{alpha}).
It is left to impose the constraint $\sum_i\delta n_i=\delta n$ in order to set the value of $\gamma$. This yields
\begin{equation}\label{222}
\frac{\gamma}{2}=\frac{\delta n}{\bar{n}\alpha(\rho_0)}.
\end{equation}
Equations~(\ref{111}) and (\ref{222}) yield Eqs. (\ref{optn}) and~(\ref{optlam}).

\section{Optimal density and flux fields for the RWs}\label{qrws}

For the non-interacting RWs one can solve Eq.~(\ref{pois}) explicitly. Since both $\phi_i$ and $\bar{\rho}_{\text{RWs}}$ are harmonic functions [see Eq.~(\ref{phirw})], we can rewrite the r.h.s. of Eq.~(\ref{pois}) as
$$
\nabla\cdot\left[\sigma(\bar{\rho})\nabla\phi_i\right]=2D_0\nabla\cdot\left(\bar{\rho}_{\text{RWs}}\nabla\phi_i\right)=D_0\nabla^2\left(\bar{\rho}_{\text{RWs}}\phi_i\right).
$$
Therefore, Eq.~(\ref{pois})  can be rewritten as a Laplace's equation:
\begin{equation}
\nabla^2\left[D_0\rho_{1_{\text{RWs}}}-D_0\bar{\rho}_{\text{RWs}}\sum_i\lambda_i\phi_i\right]=0.\label{poisrw}
\end{equation}
By virtue of the boundary conditions for $\bar{\rho}$, $\phi_i$ and $\rho_1$, the function under the Laplacian vanishes on all the patches and at infinity, whereas its normal derivative vanishes on the reflecting part of the boundary $\Omega_r$. As a result,  this function vanishes everywhere, and we obtain
\begin{equation}
\rho_{1_{\text{RWs}}}(\mathbf{x})=\bar{\rho}_{\text{RWs}}(\mathbf{x})\sum_i\lambda_i\phi_i(\mathbf{x}).\label{q1rw}
\end{equation}
Now we use Eq.~(\ref{sigmalin}) alongside with $C_{ij}=\delta_{i,j}\bar{n}_i$ (see Sec. \ref{covv}) and obtain:
\begin{equation}\label{lamrw}
\lambda_i=\frac{\delta n_i}{\bar{n}_i}.
\end{equation}
Plugging this relation into Eq.~(\ref{q1rw}) we arrive at Eq.~(\ref{qrw}). Also, using Eq.~(\ref{q1rw}), we can evaluate the flux deviation
$$
\boldsymbol{\delta J}=-\nabla\left[D(\bar{\rho})\rho_1\right]+\sigma(\bar{\rho})\nabla p_1.
$$
Substituting here $p_1$ from Eq.~(\ref{p1}) and using (\ref{lamrw}), we arrive at Eq.~(\ref{jrw}).

\section{Optimal profiles for the total absorption current}\label{totqp}
We start from $p_1$ from Eq.~(\ref{p1}). All $\lambda_i$ are given by Eq.~(\ref{optlam}). Then, using Eqs.~(\ref{phii}) and (\ref{alpha}), we obtain:
\begin{equation}
p_1=\frac{\delta n}{\bar{n}}\frac{I_1(\rho_0)}{I_2(\rho_0)}\tilde{I}_1(\bar{\rho},\rho_0),
\end{equation}
where $\tilde{I}_1(\bar{\rho},\rho_0)$ is defined by Eq.~(\ref{itild}).
Now we turn to Eq.~(\ref{pois}) for $\rho_1$.  Using
\begin{equation}
\sigma(\bar{\rho})\nabla\tilde{I}_1(\bar{\rho},\rho_0)=\nabla\tilde{I}_2(\bar{\rho},\rho_0),
\end{equation}
we obtain
\begin{eqnarray}
\sigma(\bar{\rho})\nabla p_1=\nabla\left[\frac{\delta n}{\bar{n}}\frac{I_1(\rho_0)\tilde{I}_2(\bar{\rho},\rho_0)}{I_2(\rho_0)}\right].
\end{eqnarray}
As a result, Eq.~(\ref{pois}) for $\rho_1$  becomes Laplace's equation,
\begin{equation}
\nabla^2\left[D(\bar{\rho})\rho_1-\frac{\delta n}{\bar{n}}\frac{I_1(\rho_0)\tilde{I}_2(\bar{\rho},\rho_0)}{I_2(\rho_0)}\right]=0,
\end{equation}
with inhomogeneous boundary conditions on the absorbing patches [as can be deduced from the boundary conditions (\ref{dbcgen1})-(\ref{dbcgen14}) for $\rho_1$ and the definition (\ref{itild})  of $\tilde{I}_2$].
The solution gives $\rho_1$ as a function of the average density field   $\bar{\rho}(\mathbf{x})$ in Eq.~(\ref{qq}).


\begin{thebibliography} {99}
\bibitem{blayers} In analogy with other problems of this type \cite{b,fullabsorb,surv}, we expect that the full time-dependent solution
of the MFT problem exhibits
narrow boundary layers in time near $t=0$ and $t=T$ which accommodate the boundary conditions~(\ref{bt0}) and (\ref{bt1}). These boundary layers
give only a subleading contribution to the action (\ref{actionmain}) which we do not attempt to calculate.
\end{thebibliography}
\end{document}